# Environmental impacts and monetary costs of healthy diets worldwide

*Yan Bai[1*], Elena M. Martinez[2*], Mizuki Yamanaka[1], Marko Rissanen[1], Anna Herforth[3], William A. Masters[2,4]*

Manuscript accepted for publication at *Nature Food*, last revised 3 Nov. 2025

[*] First authors, contributed equally

1. Development Data Group, The World Bank, Washington, DC, USA
2. Friedman School of Nutrition Science and Policy, Tufts University, Boston, MA, USA
3. Food Prices for Nutrition project, Tufts University, Boston, MA, USA
4. Department of Economics, Tufts University, Medford, MA, USA

Correspondence to Y. Bai (ybai@worldbank.org), E.M. Martinez (elena.martinez@tufts.edu), and W.A. Masters (william.masters@tufts.edu)

**Abstract**

Using real-world food price and greenhouse gas (GHG) emissions data for locally available food items in 171 countries, we measure how healthy diets could be obtained with the lowest possible emissions, compared to costs and emissions of the least expensive options and foods most commonly consumed. We find that foods with the lowest GHG emissions for a healthy diet would emit 0.67 kg $CO_2e$. A healthy diet using the least expensive items in each country would emit 1.65 kg $CO_2e$ and cost $3.68 in 2021, while using foods most commonly consumed would emit 2.44 kg $CO_2e$ and cost $9.96. Ninety-one percent of the difference in emissions between the lowest-cost and lowest-emissions diets is driven by animal-source foods and starchy staples. Other food groups, especially fruits and vegetables, vary widely in cost but not in emissions. Results show how changes in food policy and choice can most cost-effectively support healthier and more sustainable diets worldwide.

**Funding**

This work was conducted as part of the Food Prices for Nutrition project (INV-016158) and the Innovative Methods and Metrics for Agriculture and Nutrition Actions project (INV-002962), both funded by the Gates Foundation and UK International Development as detailed here: https://sites.tufts.edu/foodpricesfornutrition. We are grateful to all project collaborators for their contributions to this collective effort.

## Environmental impacts and monetary costs of healthy diets worldwide

Since 2020, the Cost and Affordability of a Healthy Diet (CoAHD) has been adopted by FAO[1] and the World Bank[2] for global food security monitoring[25], and used within countries to track subnational variation over time and space[3]. This work reveals that about 2.6 billion people worldwide cannot afford the least expensive locally available items in sufficient quantities from each food group required for health[4]. CoAHD is a diagnostic indicator, revealing the degree to which poor diet quality might be caused by limited supply of low-cost options in each food group, insufficient income available for food, or displacement of healthy diets by other foods[5]. All three factors influence malnutrition to varying degrees[6], and the usefulness of CoAHD has led to its adoption by a variety of national government agencies, international organizations and research institutions to guide food policies[7] and nutrition interventions[8] towards availability, access, and consumption of healthy diets worldwide[9].

Identifying low-cost options for a healthy diet is an essential step towards allowing all people to meet their nutritional needs. This study extends prior work on foods with the lowest monetary cost to consider foods with the lowest greenhouse gas (GHG) emissions, which is increasingly important[10,11,12]. Prior work has shown that within nutritionally similar food groups, lower-cost items generally have lower GHG emissions[13], so healthy diets that follow dietary guidelines and are composed of lower-cost options tend to be more environmentally sustainable than the higher-cost diets commonly consumed in higher-income settings. Prior work has also shown that the least-cost diets identified to compute the CoAHD have comparable GHG emissions and water use to the EAT-Lancet diet, a diet pattern developed with sustainability targets[14,18]. However, prior studies have not yet quantified the lowest possible emissions following the CoAHD framework, and what type of substitutions would have the greatest reductions in GHG per unit of cost.

This study measures the frontier of the lowest possible GHG emissions for a healthy diet, defined as the minimum $CO_2$-equivalent emissions per day required to meet global Healthy Diet Basket (HDB) targets[14]. We compare this benchmark to the emissions and cost of healthy diets using the least expensive and then the most commonly consumed items in each country[12,13]. The HDB is an international dietary standard introduced in 2022 to define the types and quantities of food groups needed for an active and healthy life, based on commonalities among national dietary guidelines[15]. HDB targets serve as the foundation of measuring the Cost of a Healthy Diet (CoHD), which tracks economic access to sufficiently nutritious food using retail price data. In this study we quantify the emissions and cost of different ways to meet HDB targets using foods commonly sold around the world, revealing the range of possibilities through which the existing global food system can meet human nutritional requirements.

Our analysis uses the 2021 cycle of International Comparison Program (ICP) data, providing the most recent global coverage of item availability and price in almost every country of the world[16]. To quantify the role of food choice in diet cost and emissions, we compare five benchmark item selection rules, choosing foods with each country's (1) lowest monetary cost, (2) lowest GHG emissions, (3) most commonly consumed items, (4) all commonly consumed items weighted by that product's share of consumption, and (5) all commonly consumed items equally weighted.

The five diets we construct for this study all adhere to the same nutritional standard set by the global reference HDB, thereby revealing trade-offs between affordability, sustainability and other criteria when meeting human health needs as specified by UN agencies[17] in ways that can be compared to national dietary guidelines and private initiatives such as the EAT-Lancet Commission targets[18,19]. Diets 1 and 2 are benchmarks showing the frontier of what is possible: diet 1 uses the least expensive locally available items in each food group, while diet 2 uses the items with lowest GHG emissions. Diets 3 and 4 are based on actual consumption patterns within food groups, showing the most realistic way to meet Healthy Diet Basket targets under current conditions in each country: diet 3 uses the products that are most consumed in each country, while diet 4 uses all products in proportion to actual consumption as identified by food balance sheets. Diet 5 uses those same products in equal proportions to show the full range of possibilities. Both diets 4 and 5 include all retail items reported to be available for each Healthy Diet Basket food group, using Monte Carlo sampling to capture the full distribution of costs and GHG emissions of locally available foods in each country.

## Results

**Figure 1** shows kernel density plots of healthy diet costs (Panel A) and GHG emissions per day (Panel B), calculated using retail food prices and availability for 171 countries in 2021. These five diet scenarios reflect different food selection strategies, ranging from idealized benchmarks to the consumption patterns actually observed in each country. All diets meet the Healthy Diet Basket (HDB) requirements, consisting of 11 items from six food groups.

From the distributions shown in **Panel A** of Figure 1, the mean cost of the least expensive healthy diet was \$3.68 (SD = 0.75, IQR: 3.19-4.02) in 2021, when all prices are expressed in purchasing power parity terms at 2021 prices. Switching to the food items with the lowest emissions (diet 2) and the most commonly consumed food in each country, identified by food balance sheets (diet 3) would nearly double the mean cost to \$6.95 (SD = 1.86, IQR: 5.51-8.02) and \$6.33 (SD = 1.91, IQR: 4.96-7.15). Using the full range of foods weighted by consumption patterns as identified by food balance sheets (diet 4) leads to much higher mean cost of \$9.96 (SD = 4.92, IQR: 6.98-11.46), while randomly selected foods (all items with equal weight) (diet 5) would have even higher mean cost of \$13.74 (SD = 5.52, IQR: 10.05-16.09) in 2021. As shown in the supplemental information, t-tests confirmed that all cost differences from diet 1 were significant at the 0.05 level (Suppl. Table 1).

From the emissions data in **Panel B** of Figure 1, the GHG emissions from least-cost diets (diet 1) were modest, at a global mean of 1.65 kg $CO_2e$ (SD = 0.56, IQR: 1.24-2.07), which is roughly half the estimated global mean of 3.24 kg $CO_2e$ per person per day for current diets[14]. Least-cost diets based on the most commonly consumed foods (diet 3) averaged 1.86 kg $CO_2e$ (SD = 0.61, IQR: 1.34-2.28), while the lowest emissions for a healthy diet (diet 2) were substantially lower at a mean of 0.67 kg $CO_2e$ (SD = 0.10, IQR: 0.61-0.71). Using the full range of foods weighted by consumption patterns in each country (diet 4) would cause considerably higher emissions, with mean values of 2.44 kg $CO_2e$ (SD = 1.27, IQR: 1.52-2.96), while equal weight across all items (diet 5) increases emissions further, to a mean of 3.03 kg $CO_2e$ (SD = 1.50, IQR: 1.89-3.92) (Suppl. Table 2). Monte Carlo results in the supplemental information show 95% confidence intervals for all countries and years (Suppl. Tables 3 and 4), revealing how selecting lower-priced items for a healthy diet generally, but not always, leads to lower emissions (Suppl. Fig. 1).

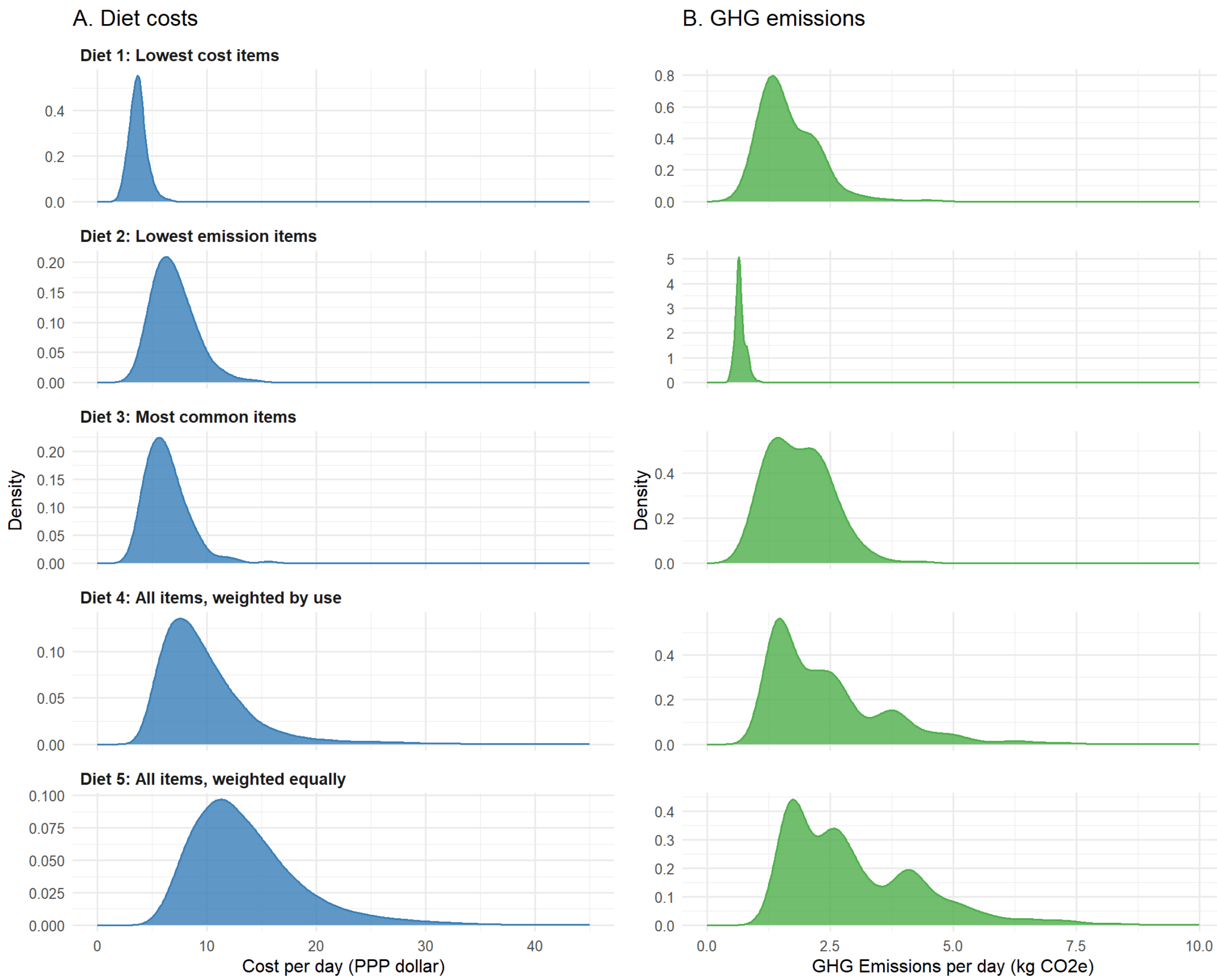

**Figure 1 | Distribution of monetary costs and greenhouse gas emissions for five healthy diets, 2021.** (A) Monetary costs (in PPP dollars per day) are shown for five dietary scenarios: Diet 1 (lowest cost items), diet 2 (lowest emission items), diet 3 (most common items), diet 4 (all items weighted by use), and diet 5 (all items weighted equally). (B) Greenhouse gas emissions (kg $CO_2e$ per day) are shown for the same five diets. The distributions are based on estimates for 171 countries for diets 1, 2, and 5, and 159 countries for diets 3 and 4. Data for diets 4 and 5 are Monte Carlo results with 1,000 iterations for each country. Because these charts display density distributions, the absolute height of each curve reflects a scaling factor rather than an intrinsic value; using different y-axes improves interpretability. Underlying data sources include national annual average food prices from the International Comparison Program (ICP), food consumption patterns from the Food Balance Sheets (FBS), and GHG emissions from Petersson et al. (2021). Further methodological details are provided in the Methods and Data sections.

**Figure 2** shows variation in costs and emissions per day within each of the six food groups needed for a healthy diet. Across all diet scenarios, animal-source foods are by far the largest contributors to GHG emissions and are typically also the highest cost food group.

Focusing on costs shown in **Panel A**, when choosing only the least expensive items (diet 1), the mean daily cost for animal-source foods was \$1.00 (SD = 0.28, IQR: 0.80-1.19), the mean cost of vegetables was \$0.77 (SD = 0.30, IQR: 0.54-0.95), the mean cost of fruits was \$0.77 (SD=0.29, IQR: 0.60-0.88), and the mean cost of starchy staples was \$0.58 (SD=0.20, IQR: 0.46-0.68). When selecting the lowest-emission items (diet 2) or the most commonly consumed items (diet 3) compared to the lowest-cost items (diet 1), the increase in cost varies but it is more substantial for some food groups. Specifically, animal-source foods cost \$2.52 (SD=1.04, IQR: 1.84-2.91) for diet 2 (>150% higher than least-cost items) and \$1.54 (SD = 0.59, IQR: 1.13-1.86) for diet 3 (>50% higher than least-cost items), reflecting the added expense of minimizing GHG emissions or meeting consumer preferences. Monte Carlo sampling from all available foods with actual consumption patterns shows the presence of higher-cost options in diet 4, with equal weighting to even higher costs in diet 5. Supplemental data provide additional insights into these diets in Suppl. Table 5.

Focusing on emissions shown in **Panel B**, animal-source foods have both the highest levels and the greatest variation in GHG emissions per day. When choosing the least expensive items in each country (diet 1), mean daily GHG emissions for animal-source foods were 0.97 kg $CO_2e$ (SD = 0.58, IQR: 0.59-1.43) in 2021. Selecting low-GHG food items reduces emissions in most food groups but by far the largest reductions in emissions are from animal-source foods and starchy staples, which account for 91% of the reduction in emissions between the lowest-cost (diet 1) and lowest emissions (diet 2) scenarios. The relatively high contribution from starchy staples compared to other food groups reflects the large daily recommended intake of starchy staples required to meet national dietary guidelines and the global HDB targets (Suppl. Table 6), rather than high emissions per kilogram. In contrast, the total emissions and variation in emissions from fruits, vegetables, legumes, and oils were minimal (Figure 2B). Shifting to lower-emission items within these food groups would have a limited effect on total diet-related emissions. However, items in these food groups vary widely in price, so shifting to lower emissions items could increase diet cost significantly. For example, fruits cost \$0.77 (SD = 0.29, IQR: 0.60-0.88) for diet 1 and \$1.57 (SD = 0.79, IQR: 1.03-1.87) for diet 2, accounting for 24% of the difference in cost between these two diet scenarios. However, fruits accounted for only 5% of the reduction in emissions between the lowest cost (diet 1) and lowest emissions (diet 2) scenarios. A diet composed of the lowest-emissions starchy staples and animal-source foods and lowest-cost items in other food groups would cost \$5.77 on average and have just 0.1 kg $CO_2e$ higher emissions than the lowest-emissions diet.

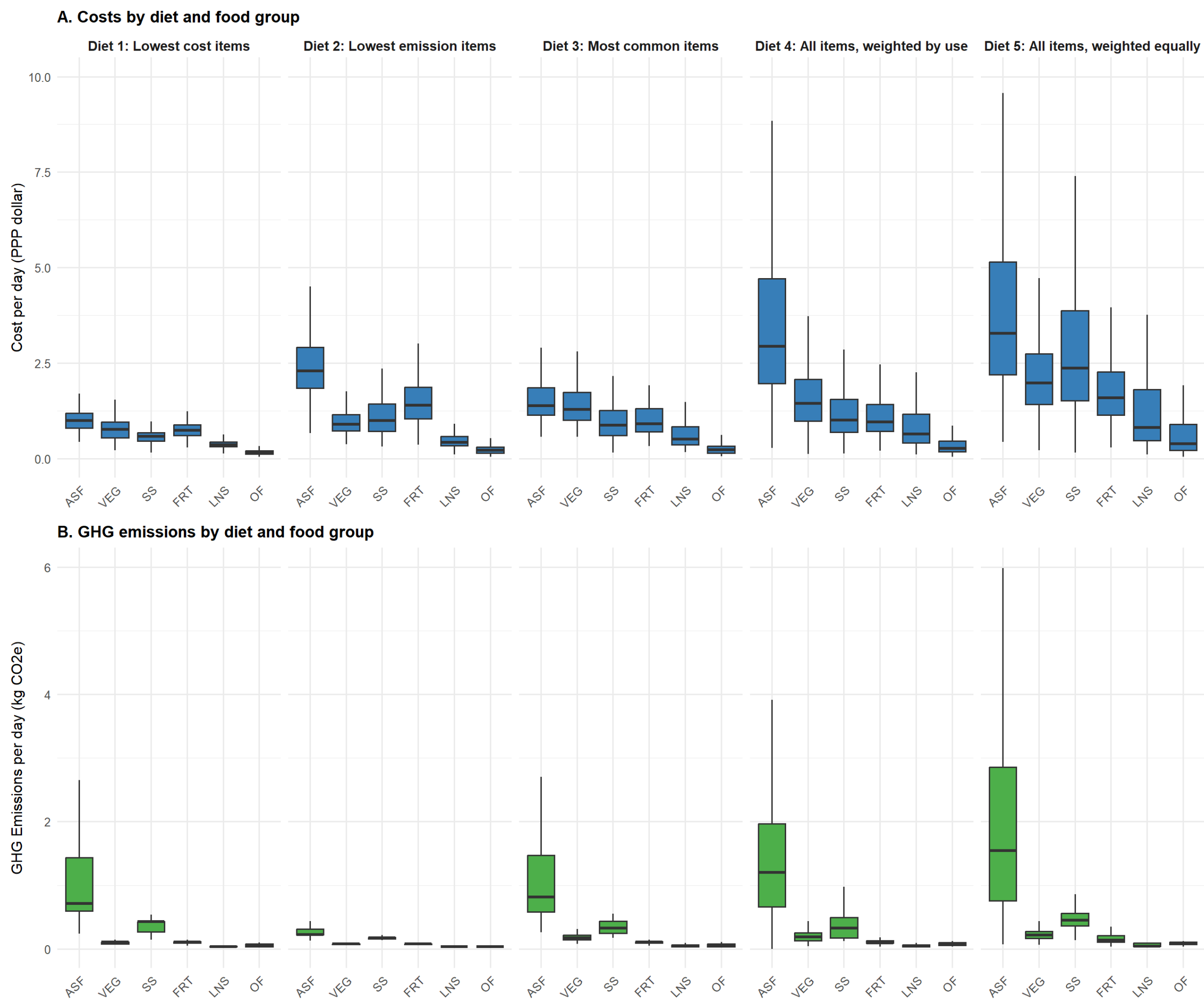

**Figure 2 | Distribution of cost and emissions by food group for five healthy diets, 2021.** (A) Costs by diet and food group: Monetary costs (in PPP dollars) are shown for each food group in five dietary scenarios: Diet 1 (lowest cost items), diet 2 (lowest-emission items), diet 3 (most common items), diet 4 (all items weighted by use), and diet 5 (all items weighted equally). (B) Greenhouse gas emissions by diet and food group: Greenhouse gas emissions (kg $CO_2e$ per day) are shown for the same diets. Food groups are specified in the Healthy Diet Basket, as 300 kcals of animal-source foods (ASF), 110 kcals of vegetables (VEG), 1,160 kcals of starchy staples (SS), 160 kcals of fruits (FRT), 300 kcals of legumes, nuts, and seeds (LNS), and 300 kcals of oils and fats (OF). Retail food prices are national averages in 2021 from the International Comparison Program, and emissions data are from Petersson et al. (2021). Box plots represent the interquartile range (IQR) of costs or emissions across 171 countries for diets 1, 2, and 5, and 159 countries for diets 3 and 4, with median values shown as horizontal lines. Whiskers extend to 1.5 times the IQR or the highest and lowest values within that range.

**Figure 3** further shows the actual range of monetary cost and greenhouse gas emissions of items within each food group needed for a healthy diet. The two columns use identical data, shown along a linear scale in the left column to reveal variation at higher levels of cost and emissions, and a log scale in the right column to reveal variation at lower levels of cost and emissions. For visual clarity, these results omit diet 5, which continues the progression from diets 2, 3 and 4.

As revealed in **Figure 3**, the available options for healthy diets have a range of emissions at each price level, especially for animal source foods and starchy staples. When choosing only the lowest cost foods in each country (diet 1), the range of emissions at each price level narrows considerably but remains wider for animal source foods and starchy staples compared to other food groups. When choosing the lowest emissions foods (diet 2), the range of monetary cost for each food group widens. For animal source foods and starchy staples, the lowest emissions items have higher monetary cost yet considerably lower emissions compared to lowest cost items. For other food groups, lowest emissions options have higher monetary cost but only very marginally lower emissions. When choosing the most commonly consumed items (diet 3), all food groups include a variety of higher-emission items. The scatterplots for diets 3 and 4 also show a positive correlation between greenhouse gas emissions and monetary cost per day for commonly consumed items in and between food groups, revealing how the additional resources used to produce more valuable types of food often incur additional emissions as well as additional monetary cost.

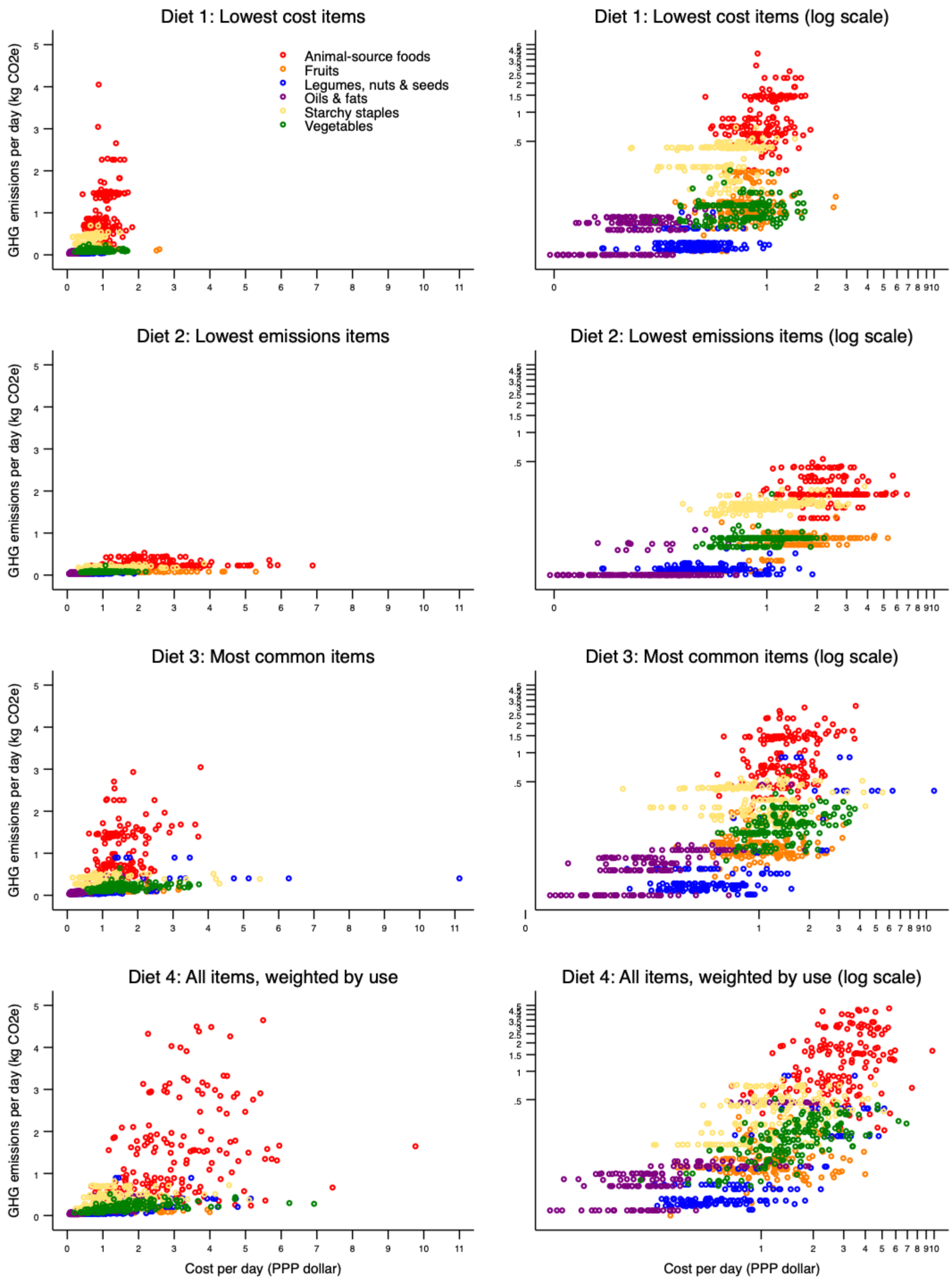

**Figure 3 | Greenhouse gas emissions and cost per day of items in each food group for four healthy diets, 2021.** Both columns show the same data, in linear scale (on the left) and natural log scale (on the right), with each food group's emissions (in kg of $CO_2e$) and monetary cost (in PPP dollars) per person per day, for 165 countries with diets 1 and 2, and 157 countries with diets 3 and 4. Dots represent each country, using the median cost selections of the Monte Carlo results for diet 4. Emissions and costs are color-coded for each of six Healthy Diet Basket (HDB) food groups: animal-source foods, starchy staples, legumes, nuts and seeds, vegetables, fruits, and oils and fats.

**Figure 4** uses a mosaic plot with rectangles proportional to dietary energy and GHG emissions from each item, color-coded by food group, in the global average of three benchmark healthy diets using only the lowest cost items (diet 1), the lowest emissions items (diet 2), and the most commonly consumed items (diet 3). Some items make large energy contributions in all three diet scenarios, indicating that they are consistently inexpensive, commonly consumed, and have low emissions, notably wheat, maize, white beans, apples, onions, and carrots. Other foods are inexpensive and commonly consumed but have higher emissions than other options in the same food group, such as rice, pasta, palm oil, milk, chicken, and beef. A third category of food items, such as oats and sardines, often have the lowest emissions in their food group but are neither the least expensive nor the most commonly consumed options. As shown in Figures 1 and 2, the lowest emission items generally also have lower cost per day than other commonly consumed items in their food group. The largest tradeoffs shown in Figure 4 are for starchy staples, because rice is inexpensive and widely consumed in large quantities with higher emissions (averaging 2.19 kg $CO_2e$) compared to wheat (0.57 kg $CO_2e$) and maize (0.48 kg $CO_2e$). In contrast, oats have relatively low average emissions (0.67 kg $CO_2e$) but are almost never either the lowest cost or the most widely consumed starchy staple.

Item selection is based only on food group targets specified in the Healthy Diet Basket derived from national dietary guidelines, so the resulting mix of foods also generally meets all essential nutrient requirements as measured by Mean Adequacy Ratios (MAR) around 0.95 across diets. Minor shortfalls are observed for calcium, vitamin C, riboflavin, and vitamin B12 in diet 1; vitamin C, riboflavin, and calcium in diet 2; and calcium, riboflavin, vitamin B12, and vitamin A in diet 3. These gaps may partly reflect the use of national average food price data and nutrient composition values primarily from the USDA database, but they highlight nutrients that could warrant attention when designing cost-effective, low-emission diets (Suppl. Table 7).

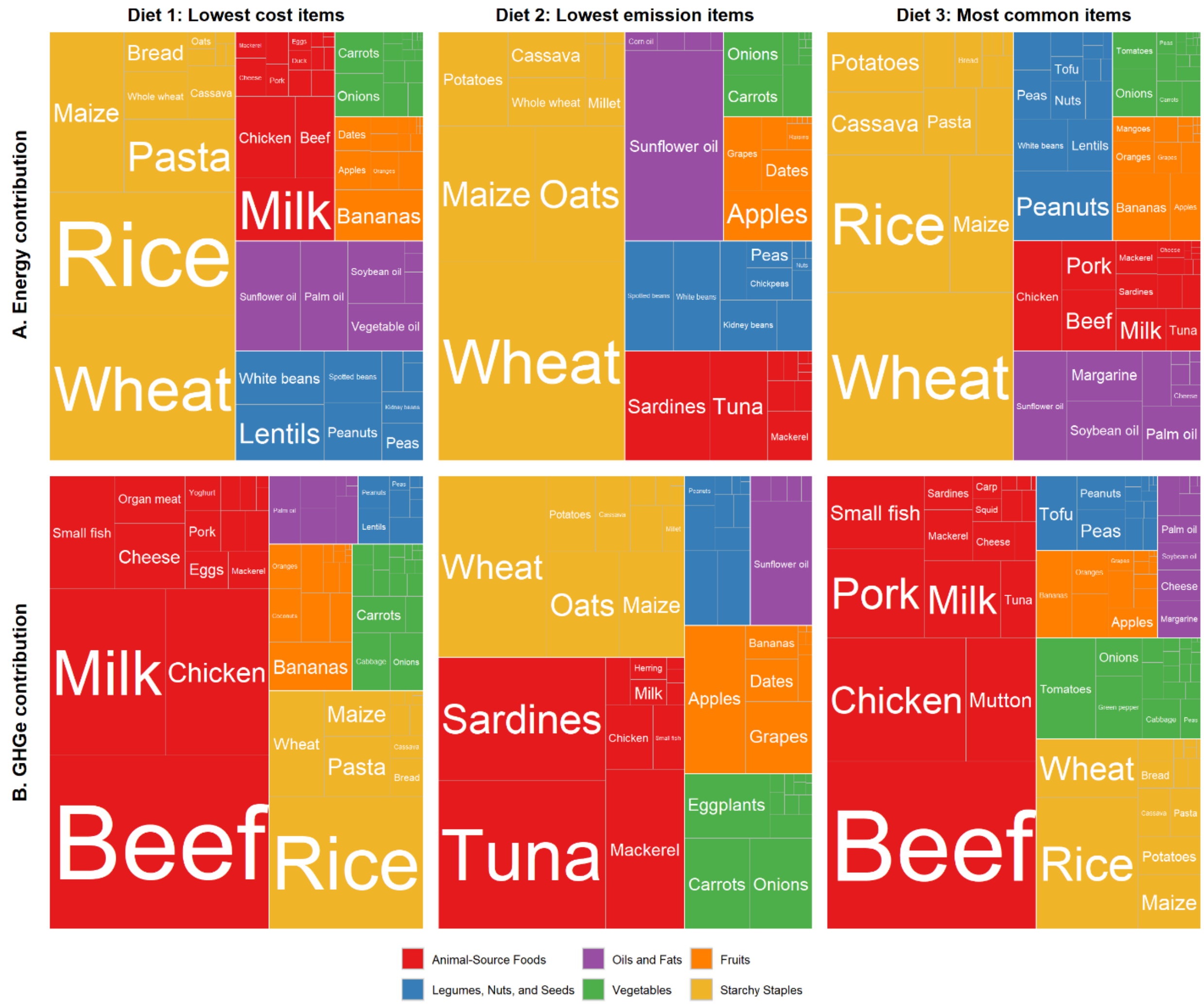

**Figure 4 | Shares of dietary energy and GHG emissions by item in three benchmark heathy diets, 2021.** (A) Dietary energy shares of food items are color-coded in each of six Healthy Diet Basket (HDB) food groups: animal-source foods, starchy staples, legumes, nuts and seeds, vegetables, fruits, and oils and fats. (B) Shares of greenhouse gas emissions are shown for each item in the same food groups. Both sets of results are shown for three extreme benchmark diets scenarios: Diet 1 (lowest cost items), diet 2 (lowest-emission items), and diet 3 (most common items) in each country. Data shown are from 171 countries for diets 1 and 2, and 159 countries for diet 3.

**Figure 5** uses nonparametric curves to test whether diet costs vary systematically with a country's national income. All diet costs are expressed in real terms, converted using Purchasing Power Parities (PPPs) for household consumption expenditures to account for price level differences across countries. These healthy diet costs differ greatly from actual food spending, which rises with national income due to use of more expensive items within food groups and obtaining larger shares of dietary energy from relatively expensive food groups, especially animal-source foods. The cost of the five benchmark healthy diets does not rise with national income, and some diets are significantly less expensive in higher-income countries. A similar analysis for GHG emissions shows no clear patterns across diets (Suppl. Fig. 2).

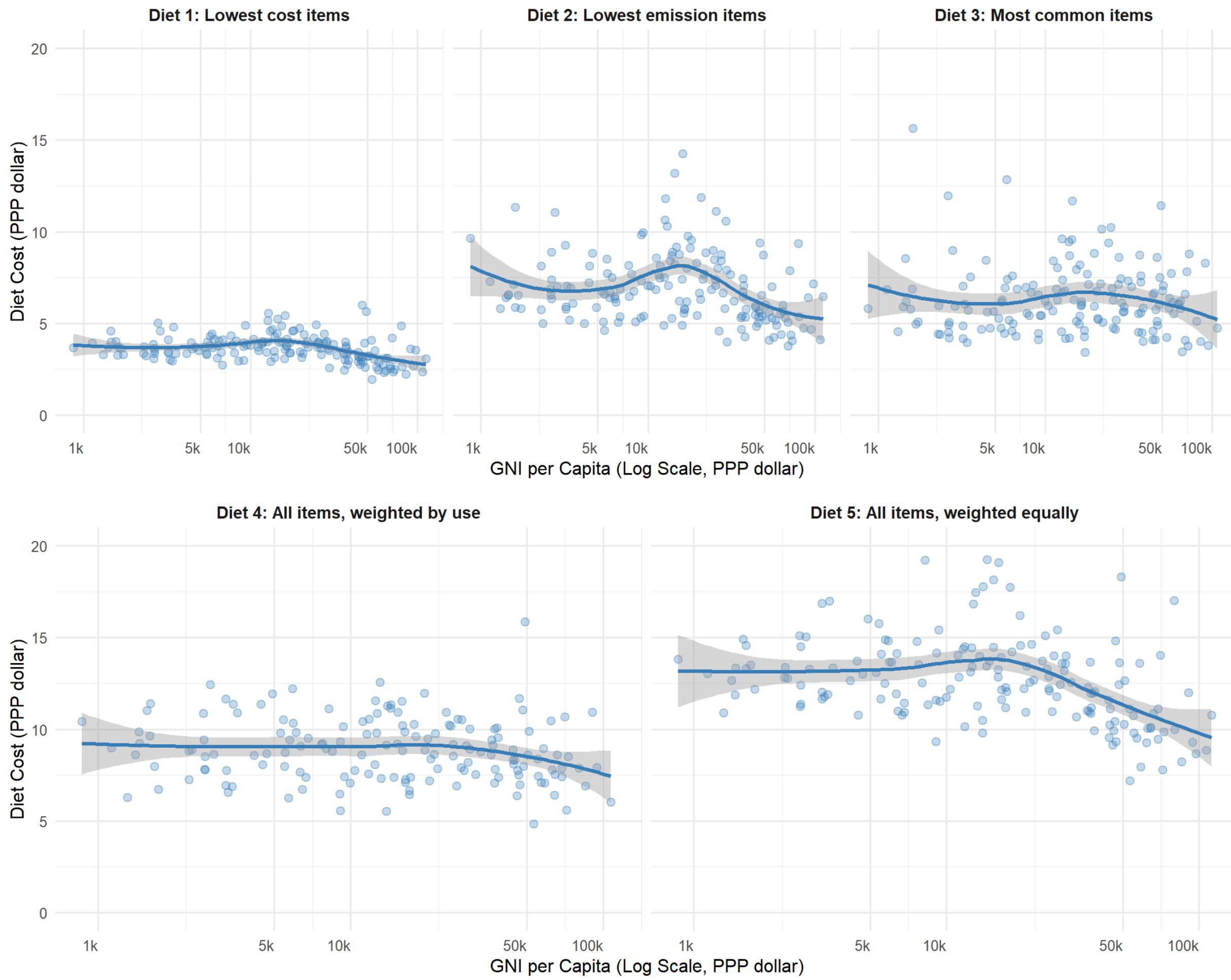

**Figure 5 | Total cost per day by level of national income for five healthy diets, 2021.** Lines show the mean and 95% confidence interval of monetary cost at each level of Gross National Income (GNI), both in PPP dollars per person per day, for 165 countries with diets 1, 2, and 5, and 157 countries with diets 3 and 4. Dots represent each country, using the median of the Monte Carlo results for diets 4 and 5. Lines estimated by LOESS regression to test for systematic variation in diet costs associated with national income.

In low-income countries, a large share of the population cannot afford a healthy diet. The geographic distribution of unaffordability is shown in **Suppl. Fig. 3**, which uses color-coded maps to show the percentage of the population unable to afford healthy diets in 2021 by country, when selecting the lowest cost locally available food items in each country in **Panel A,** the lowest emission items in **Panel B**, and the most commonly consumed items in **Panel C**. The highest concentration of unaffordability is in the lower-income countries of Sub-Saharan Africa and South Asia where over 75% of people are unable to afford any of these benchmark healthy diets, whereas less than 10% of people are unable to afford these diets in higher-income countries. In middle income countries such as China, Brazil, Mexico, Argentina, and Russia, fewer than half of the population are unable to afford healthy diets, even when switching from the least expensive items to the most commonly consumed or least emitting items. Populations for whom these benchmark diets are affordable often consume other foods instead, for reasons such as taste and preferences, time use and cooking costs, or aspirations shaped by food culture and marketing.

## Discussion

We estimate the greenhouse gas emissions and monetary cost of foods needed to meet a global standard for healthy diets, using locally available items in 171 countries around the world to compare five criteria for item selection in each country. Our first benchmark uses only the least expensive locally available items, which is the criterion used internationally by UN agencies, the World Bank, and country governments for global monitoring of food access based on the monetary cost and affordability of healthy diets. We then compare the cost and GHG emissions of lowest-cost diets to four other options: (a) the lowest GHG emissions items available in each country, (b) the most commonly consumed items in each country, (c) all items selected in proportion to use, and (d) all items selected randomly.

Our principal finding is that differences in the cost and GHG emissions between these four diet scenarios are due primarily to choices within just two food groups: animal source foods and starchy staples. Within these two food groups, higher cost items usually but do not always have higher emissions. For example, among animal source foods, milk and poultry meat are often among the least costly items for a healthy diet, while the least emitting items often include small fish. Other items have higher cost and higher emissions, including commonly consumed foods such as beef. Selection among animal-source foods is important due to their wide range and high levels of both costs and emissions, while selection among starchy staples is important due to the much larger quantity needed for a healthy diet. In contrast, fruits, vegetables, oils and fats, or legumes, nuts and seeds vary more widely in terms of monetary cost than in terms of emissions. These findings extend prior work that focuses on the monetary cost and affordability of healthy diets[3,4], and complement prior work that focuses on shifts between food groups[18,19,20].

The alternatives compared in this study reveal how the foods needed for health differ in terms of both environmental impacts and monetary cost. These results underscore the importance of reducing emissions from the animal source foods and starchy staples needed in a healthy diet, in addition to reducing the monetary cost of the most affordable options for improved access to all required food groups, including vegetables and fruits that are relatively expensive. Actual choices depend not only on price and income but also other factors such as taste and preferences, time use and cooking costs, or habits and aspirations. In each country, demand, local production, trade patterns, and government policies and programs all influence which foods are consumed most frequently. Our comparisons between the least emitting and least cost options to commonly consumed diets reveal how locally available foods in each country could support shifts toward healthier and more sustainable diets that are also more affordable.

The role of food choice is particularly important for cost and sustainability when selecting among animal-source foods, and to a lesser extent starchy staples. Other plant-source foods are already low in emissions but vary widely in cost. In response to climate change, many government agencies, international organizations, and civil society groups have sought to incorporate emission reductions into decision making. Our findings confirm that in most cases, choosing lower cost items within the animal source food group also leads to lower emissions, although the absolute least cost items may not have the absolutely lowest GHG emissions. This study shows how there are expensive, high-emissions food items in all countries that are substitutable, revealing opportunities to meet health needs in more sustainable and affordable ways. The methods shown here can be used to guide policies and investments towards more sustainable as well as healthier diets at each location around the world.

Limitations of the study include the data on availability and price of commonly used foods in each country. At any market location there are often thousands of distinct items for sale, each from a variety of origins being sold at different prices over time. For this study we use availability and annual average prices for the most commonly sold items in up to 173 countries, accounting for over 97% of the global population (Suppl. Table 8). These ICP data are the only source of retail food prices and availability with worldwide coverage, but by design the data are limited to standardized items sold in multiple countries. This includes foods sold in just a few countries such as amaranth leaves, but omits items that are widely consumed in only one country such as teff in Ethiopia. Future work looking within countries could use the even more detailed market data, capturing seasonal and spatial variation in emissions as previously done only for diet costs. For this baseline study, observed national average prices in 2021 are static and do not respond to dietary shifts. If widespread dietary shifts were to occur, for example due to a change in government policies, prices would adjust, potentially changing which products are the least expensive options in each food group. Future work could model specific scenarios using partial or general equilibrium frameworks that track how consumers and producers might adjust to each change.

A second limitation of the study concerns the data on GHG emissions, which could vary depending on how each food is produced and distributed to consumers at each place and time. For this study we use the most recent and complete peer-reviewed compilation global average estimates of GHG emissions, based on studies from 78 countries on 6 continents, all harmonized to represent emissions from the farm to point of retail sale. Country-level estimates of the GHG emissions of many of the food items in our analysis are not currently available. Future work could incorporate more granular estimates of the climate footprints that may be produced, transported, processed and stored in different ways prior to consumption, potentially extended to other environmental or social impacts as in other work on sustainable diets.

A third limitation is methodological, regarding item selection criteria other than the least expensive or least emitting food choices. Observed diets depend on a wide range of factors driving the consumption of each type of food, leading to different dietary patterns for each person over time. In this study, we use national average food supplies as measured by food balance sheets, which infer each country's total consumption per person of each product category from a country's total production plus imports minus exports, nonfood uses, and losses to the point of retail sale. The consumption patterns revealed by these data are a result of the interaction of demand and supply with trade opportunities and government policies. Foods actually consumed are not necessarily those most preferred by taste or aspiration, as choices from the available options are constrained by available income and prices as well as convenience and time constraints, cooking methods and other factors. Future work could address changes in availability and price as well as food choice using individual dietary intake or household consumption surveys, including differences in kitchen and plate waste, to understand variation in what and how food is consumed.

## Methods

### *Data*

This study combines three kinds of data about each food item: (1) availability and price in each country; (2) share of that product in the country's food supply; and (3) global average GHG emissions associated with that product. Sources for each are described below.

### 1. Food item availability and price

We use retail food prices and availability from the International Comparison Program (ICP)[14]. Our analysis of ICP data excludes discretionary items not needed in healthy diets as defined by national dietary guidelines, such as salty snacks, cured meats, and sugar-sweetened products. We also exclude infant foods, beverages, herbs and spices, condiments, culinary ingredients, mixed dishes, and items lacking food composition data. The resulting database provides national average prices for a global total of 440 distinct food items across 173 countries in 2021, with each country reporting an average of 105 items across the six food groups needed for a healthy diet (Suppl. Table 8). These ICP data are reported by each national statistical organization through the World Bank, with the explicit goal of representing as many of the world's commonly consumed foods for as much of the global population as possible. Each price is a national average for the 2021 calendar year, reported in local currency per unit sold. We then converted those data into real monetary cost per unit of dietary energy for each type of food, expressed in purchasing power parity (PPP) dollars per kilocalorie (kcal) per person per day using the following steps:

1) **Converting to standard units with food composition data**: The weight or volume or each item as sold was standardized into kilograms of edible matter and then kilocalories of dietary energy by matching ICP item descriptions to food composition tables, including USDA Standard Reference 28, the USDA Food and Nutrient Database for Dietary Studies, the West

Africa Food Composition Table (FCT), and national FCTs from Bangladesh, India, Japan, and the United Kingdom.

2) **Calculating cost per day**: Prices were converted to cost per day based on dietary energy required from each food group as specified in the Healthy Diet Basket (HDB). The HDB is derived from requirements specified in national dietary guidelines, defining a balanced diet as 11 items in 6 food groups adding up to 2,330 kcals/day, of which 1,160 kcal is from two different starchy staples, 300 kcal from any type of legumes, nuts or seeds, 110 kcal from three different vegetables, 160 kcal from two fruits, 300 kcal from two animal source foods, and 300 kcal of oils and fats.

3) **Adjusting for local price levels**: The real cost of each diet in local currency terms was converted to US dollars at purchasing power parity (PPP) prices, using each country's conversion factor for private consumption sourced from the World Bank World Development Indicators (WDI)[21].

**2. Food items' greenhouse gas emissions**

To quantify the greenhouse gas emissions associated with each diet, ICP items were matched to products listed in a comprehensive global database of food-specific GHG emissions estimates[22]. This database includes estimates of cradle-to-retail-gate emissions in kg of CO2-equivalent emissions per kg of food for 324 food items, based on data from 78 countries across six continents. Those data per kg of food were converted to emissions per kcal of edible matter using data from food composition tables.

**3. Food items most commonly consumed**

To identify the quantities of each item consumed in each country, ICP items were matched to food product categories for which national totals are reported in the FAO's Food Balance Sheets (FBS) and Supply Utilization Accounts (SUA)[23]. These datasets report food supply in kcal per day per person, by country and year. Food product categories used for the FBS and SUA were matched to ICP food items, often with multiple ICP items in each food category, and multiple food categories in each food group needed for a healthy diet.

***Item selection for benchmark diets***

Our study uses the data described above to construct five benchmark diets, each measuring a different aspect of foods available and commonly consumed around the world. All five diets meet the same global Healthy Diet Basket (HDB) targets, expressed in kcals/day for each type of food. In dietary guidelines, recommended intakes are often specified in terms of volume, weight, or servings of locally appropriate reference items across different food groups. The HDB converts commonalities among guidelines into dietary energy from the six main food groups, with diversity within some food groups specified as equal shares from multiple products. Defining a healthy diet using the HDB allows item selection to vary across all of the world's dietary practices, while maintaining global standards of nutrient adequacy and protection from diet-related diseases for an average person across all ages, sexes, and activity levels, which is similar to a representative adult woman[14,15]. Energy requirements for each food group $g$ from a set of $k_g$ food items are computed as follows:

$$Q_{g,i} = E_g/k_g, g \in \{1,\dots,6\}, i \in \{1,\dots,k_g\}$$

As described earlier, for starchy staples, $E_g$= 1,160 kcal from $k_g$= 2 different items. For vegetables, $E_g$=110 kcal and $k_g$= 3. For fruits, $E_g$=160 kcal with $k_g$= 2. For animal-source foods, $E_g$=300 kcal with $k_g$= 2. For legumes, nuts and seeds, $E_g$= 300 kcal with $k_g$= 1. For oils and fats, $E_g$= 300 kcal with $k_g$= 1.

Both $Q_{g,i}$ and $P_{g,i}$ must be strictly positive, ensuring that all selected quantities and prices are non-negative:

$$Q_{g,i} > 0 \text{ and } P_{g,i} > 0$$

After computing each of the five diets, we compute and compare both their monetary costs and GHG emissions, with subscripts for each country omitted in the equations below.

**Diet 1. Lowest cost items in each food group**

Our first benchmark is the least expensive healthy diet using locally available items in each country. This is a metric of food security designed to measure changes in a population's economic access to sufficient foods for an active and healthy life, for global monitoring and national statistics on the Cost of a Healthy Diet (CoHD). The computation of CoHD selects the least-cost items within each HDB food group by rank order optimization within each food group, such that cost per day is computed as:

$$LowestCostDiet = min \sum_{g=1}^{6} \sum_{i=1}^{k_g} (P_{g,i} \times Q_{g,i})$$

where $P_{g,i}$ refers to the price of the $i$-th item in HDB food group $g$ (USD/kcal) and $Q_{g,i}$ the energy quantity of the $i$-th item in group $g$ (kcal).

**Diet 2. Lowest greenhouse gas emission items in each food group**

An alternative benchmark is the lowest possible GHG emissions using locally available items for a healthy diet in each country. This is computed by selecting $k_g$ items with the least emissions within each HDB food group $g$. If multiple items have the same level of emissions, the least-cost item is selected. This shows the limits of how currently available foods could be combined to meet human health needs with the lowest possible level of emissions, such that cost per day is as follows:

$$LowestEmissionsDiet \quad = min \sum_{g=1}^{6} \sum_{i=1}^{k_g} (P_{g,i} \times Q_{g,i}), i \in G_g$$

where $G_g$ represents the subset of $k_g$ items in food group $g$ with the minimum GHG emissions per kilocalorie, and other terms are as before.

**Diet 3. Most commonly consumed items in each food group and in each country**

We then compare the two frontiers of lowest cost and lowest emissions to healthy diets using the food products most commonly consumed in each country, based on their local supply and demand conditions as represented by national Food Balance Sheets (FBS). FBS data refer to product categories such as rice, which we match to retail items such as a 400g bag of long-grain rice. We select the least expensive item in each product category, which represents the cost and emissions of using each country's most commonly consumed products in each food group. For example, the most commonly consumed starchy staples might be wheat and potatoes in one country, but rice and maize in another, as follows:

$$MostCommonDiet \quad = min \sum_{g=1}^{6} \sum_{i=1}^{k_g} (P_{g,i} \times Q_{g,i}), i \in S_g$$

where $S_g$ represents the subset of $k_g$ items in food group $g$ that are most commonly consumed in that country. This method builds upon previous methods for calculating a form of the Cost of a Healthy Diet reflecting food preferences from consumption data[24,25].

**Diet 4. All available items, weighted by share of actual consumption in each food group and in each country**

Each country's food system offers a wide range of items that are used in varying quantities by different people. To describe the full range of costs and emissions, we use Monte Carlo methods to compute a thousand diets for each country, obtaining a distribution of diets that represent quantities consumed by applying probability weights proportional to each country's actual food supply. Within each HDB food group $g$, $k_g$ items are selected randomly, with selection probabilities proportional to that product's share of the food group. For example, within the starchy staples category, the various options for rice, wheat, potatoes, maize, or other products will all be included, with frequencies that make total quantities align with consumption by the population as a whole. In this simulation, each diet is obtained by random selection with probability weights, as given by:

$$WeightedAverageDiet = \sum_{g=1}^{6} \sum_{i=1}^{k_g} (P_{g,i} \times Q_{g,i}), i \in R_g^w, t = 1, \ldots, N$$

where $R_g^w$ denotes the subset of $k_g$ items selected with weights proportional to their actual share of dietary energy in food group *g*. The weights for item *i* are given by:

$$w_{g,i} = FBS_{g,i}$$

where $FBS_{g,i}$ represents the calorie supply of the *i*-th product in food group *g* as reported in the FBS data. For each country and year, the random selection of $R_g$ is performed $N = 1{,}000$ times, generating a distribution of dietary costs.

**Diet 5. All available items in each food group, weighted equally**

A final type of diet uses the entire range of items being sold, with equal probability of selecting each item in each food group, again using Monte Carlo methods to draw a thousand diets for each country. For each group $g$, $k_g$ items are drawn, with equal probability weights on the items in each group and repeated draws with replacement to ensure that each country's full set of diets represents all available foods in equal proportions:

$$AllAvailableDiets = \sum_{g=1}^{6} \sum_{i=1}^{k_g} (P_{g,i} \times Q_{g,i}), i \in R_g$$

where $R_g$ denotes the subset of $k_g$ items randomly selected from the full set of available items in food group *g*. The use of equal weighting for all available items in each food group is helpful as a benchmark to compare against the baseline diet costs of least-cost options or least-emitting options, as well as the most commonly consumed items.

***Affordability measure***

The **Prevalence of Unaffordability (PUA)** indicator estimates the percentage of individuals in a population whose disposable income, net of the amount required to meet basic non-food expenditures, is insufficient to afford the least expensive healthy diet in their country. Basic non-food expenditures are calculated as the average non-food expenditure shares of low -income consumers, multiplied by the international poverty lines set by the World Bank.

For low-income and lower-middle-income countries, non-food expenditure shares are derived from the second quintile of consumers, estimated at 37% and 44%, respectively. For upper-middle-income and high-income countries, non-food expenditure shares are based on the first quintile, estimated at 54%, according to recent household survey data for 71 countries, compiled by the World Bank. The income distribution data used to calculate PUA are sourced from the World Bank's Poverty and Inequality Platform (PIP)[26] and are based on 2017 PPPs. The method is summarized in Table S2.3 of the UN agencies' official report using the CoAHD method[3], with details and provided in the FAO background paper for that report[27].

Data analysis and visualization were executed in Stata SE 18 and RStudio 2023.12.0.

## Background information

### Data availability

Individual food item descriptions, availability, and prices in each country were obtained through a confidentiality agreement with the World Bank, under the International Comparison Program (ICP) data access and archiving policy (https://www.worldbank.org/en/programs/icp/data#3). Emissions data are available from their source as published in *Scientific Data* (https://www.nature.com/articles/s41597-021-00909-8), and food use data is available from its source at FAO (https://www.fao.org/faostat/en/#data/FBS). For World Development Indicator (WDI) and Poverty and Inequality Platform (PIP) data, access is provided through the World Bank's data portals (https://databank.worldbank.org/source/world-development-indicators) and (https://pip.worldbank.org).

### Code availability

Model code for matching item descriptions to food composition and nutritional requirements for computation of diet costs is available from the Food Prices for Nutrition project (https://sites.tufts.edu/foodpricesfornutrition/tools). Code for replication of this study are available at https://doi.org/10.17605/OSF.IO/6TGS4.

### Contributions

Y.B. and E.M.M. are co-first authors of this manuscript. Y.B., E.M.M., and W.A.M conceptualized the study. Y.B. and E.M.M. designed the study, analyzed the data, and drafted the manuscript. W.A.M obtained funding for the study. Y.B., E.M.M., and W.A.M. accessed and verified the data. All authors interpreted the results, revised the manuscript for important intellectual content, and read and approved the final manuscript.

### Competing interests

The authors declare no competing interests.

## Supplementary Information

Suppl. Table 1. Summary statistics for diet costs, in PPP dollars per person per day in 2021

| Diet | Obs | Mean | SD | P25 | P50 | P75 | t-test, compared to case 1, p-value (diff ≠ 0) |
|---|---|---|---|---|---|---|---|
| 1 | 171 | 3.68 | 0.75 | 3.19 | 3.68 | 4.02 | na |
| 2 | 171 | 6.95 | 1.86 | 5.51 | 6.65 | 8.02 | 0.00 |
| 3 | 159 | 6.33 | 1.91 | 4.96 | 5.97 | 7.15 | 0.00 |
| 4 | 159,000 | 9.96 | 4.92 | 6.98 | 8.86 | 11.46 | 0.00 |
| 5 | 171,000 | 13.74 | 5.52 | 10.05 | 12.67 | 16.09 | 0.00 |

Suppl. Table 2. Summary statistics for emissions, in kilograms of CO2e per person per day in 2021

| Diet | Obs | Mean | SD | P25 | P50 | P75 | t-test, compared to case 1, p-value (diff ≠ 0) |
|---|---|---|---|---|---|---|---|
| 1 | 171 | 1.65 | 0.56 | 1.24 | 1.46 | 2.07 | na |
| 2 | 171 | 0.67 | 0.10 | 0.61 | 0.64 | 0.71 | 0.00 |
| 3 | 159 | 1.86 | 0.61 | 1.34 | 1.81 | 2.28 | 0.00 |
| 4 | 159,000 | 2.44 | 1.27 | 1.52 | 2.12 | 2.96 | 0.00 |
| 5 | 171,000 | 3.03 | 1.50 | 1.89 | 2.65 | 3.92 | 0.00 |

Suppl. Table 3. Summary statistics for monetary costs using Monte Carlo simulation in each country

| No | Country | Diet 4: All food items, weighted by use | | | Diet 5: All food items, weighted equally | | |
|---|---|---|---|---|---|---|---|
| | | Diet Cost (2021PPP) | | | Diet Cost (2021PPP) | | |
| | | P2.5 | P50 | P97.5 | P2.5 | P50 | P97.5 |
| 1 | ABW | | | | 8.46 | 13.66 | 28.30 |
| 2 | AGO | 6.60 | 9.50 | 14.11 | 6.96 | 10.93 | 18.69 |
| 3 | AIA | | | | 8.45 | 13.03 | 24.36 |
| 4 | ALB | 5.69 | 8.81 | 28.25 | 7.41 | 12.83 | 33.11 |
| 5 | ARE | 4.71 | 7.89 | 16.71 | 6.27 | 11.11 | 20.83 |
| 6 | ARG | 5.37 | 9.16 | 13.34 | 7.64 | 12.10 | 16.96 |
| 7 | ARM | 6.33 | 9.85 | 15.23 | 7.98 | 13.91 | 22.00 |
| 8 | ATG | 5.77 | 10.24 | 17.11 | 9.70 | 15.10 | 21.74 |
| 9 | AUS | 3.77 | 7.11 | 15.50 | 5.62 | 9.52 | 18.24 |
| 10 | AUT | 4.07 | 7.52 | 18.92 | 5.69 | 10.06 | 20.48 |
| 11 | AZE | 6.16 | 8.87 | 16.24 | 8.23 | 14.57 | 31.56 |
| 12 | BDI | 6.23 | 10.44 | 16.28 | 7.74 | 13.82 | 27.25 |
| 13 | BEL | 3.68 | 7.79 | 21.63 | 5.51 | 10.06 | 21.37 |
| 14 | BEN | 6.23 | 10.90 | 26.96 | 7.51 | 13.35 | 26.99 |
| 15 | BFA | 4.65 | 8.88 | 18.20 | 6.67 | 12.78 | 24.40 |
| 16 | BGD | 5.96 | 9.23 | 17.88 | 8.05 | 13.52 | 23.99 |
| 17 | BGR | 5.42 | 7.75 | 22.90 | 7.88 | 12.44 | 28.13 |
| 18 | BHR | 6.10 | 9.89 | 16.54 | 8.06 | 12.65 | 22.22 |
| 19 | BIH | 5.79 | 9.61 | 22.82 | 8.92 | 14.24 | 28.26 |
| 20 | BLR | 6.08 | 10.54 | 19.98 | 8.68 | 13.99 | 25.45 |
| 21 | BLZ | 7.38 | 10.44 | 15.23 | 7.77 | 12.83 | 21.16 |
| 22 | BMU | | | | 5.42 | 8.68 | 15.18 |
| 23 | BOL | 4.03 | 5.58 | 11.22 | 6.23 | 9.31 | 17.74 |
| 24 | BON | | | | 8.89 | 14.81 | 23.60 |
| 25 | BRA | 4.36 | 6.44 | 13.33 | 7.12 | 12.29 | 24.12 |
| 26 | BRN | | | | 10.12 | 17.01 | 31.65 |
| 27 | BTN | 6.57 | 9.79 | 15.31 | 8.72 | 13.40 | 26.49 |
| 28 | BWA | 5.03 | 7.11 | 16.02 | 7.40 | 11.61 | 20.46 |
| 29 | CAF | 6.73 | 8.99 | 14.42 | 7.31 | 13.02 | 23.43 |
| 30 | CAN | 4.22 | 7.43 | 16.29 | 6.14 | 9.80 | 18.05 |
| 31 | CHE | 4.01 | 7.89 | 19.02 | 5.16 | 9.97 | 21.04 |
| 32 | CHL | 5.20 | 7.91 | 19.37 | 7.79 | 11.78 | 24.28 |
| 33 | CHN | 6.18 | 9.49 | 46.32 | 7.30 | 12.22 | 47.75 |
| 34 | CIV | 4.72 | 7.65 | 35.06 | 6.70 | 11.93 | 33.64 |
| 35 | CMR | 5.58 | 8.67 | 24.75 | 7.14 | 13.02 | 27.70 |
| 36 | COD | 4.29 | 6.29 | 10.50 | 6.69 | 10.89 | 20.85 |
| 37 | COG | 5.97 | 9.82 | 42.40 | 8.35 | 13.67 | 41.75 |
| 38 | COL | 4.20 | 7.39 | 27.04 | 6.84 | 12.07 | 31.75 |
| 39 | COM | 7.66 | 11.37 | 40.39 | 10.17 | 17.00 | 44.38 |
| 40 | CPV | 4.42 | 7.38 | 11.29 | 6.78 | 10.76 | 15.88 |
| 41 | CRI | 4.41 | 7.84 | 28.25 | 7.88 | 12.67 | 31.53 |
| 42 | CUW | | | | 7.58 | 12.21 | 22.27 |
| 43 | CYM | | | | 6.48 | 11.12 | 17.30 |
| 44 | CYP | 4.74 | 8.10 | 20.10 | 6.22 | 10.25 | 21.94 |
| 45 | CZE | 4.29 | 7.51 | 21.84 | 6.54 | 10.44 | 22.58 |
| 46 | DEU | 3.92 | 6.42 | 23.15 | 5.75 | 10.10 | 25.10 |
| 47 | DJI | 6.73 | 10.34 | 32.26 | 8.10 | 13.65 | 34.54 |
| 48 | DMA | 7.43 | 11.33 | 20.45 | 11.33 | 19.26 | 31.92 |
| 49 | DNK | 3.97 | 8.50 | 19.55 | 5.13 | 9.85 | 19.60 |
| 50 | DOM | 5.27 | 8.38 | 14.15 | 8.02 | 13.45 | 26.17 |
| 51 | DZA | 5.56 | 11.16 | 45.49 | 7.92 | 13.43 | 47.66 |
| 52 | ECU | 4.75 | 7.31 | 13.91 | 6.28 | 11.28 | 19.40 |
| 53 | EGY | 5.56 | 9.19 | 30.95 | 7.78 | 14.48 | 35.79 |
| 54 | ESP | 4.70 | 8.89 | 22.84 | 5.63 | 9.56 | 21.61 |
| 55 | EST | 4.67 | 8.31 | 21.64 | 6.59 | 11.58 | 24.30 |
| 56 | ETH | 6.93 | 10.86 | 17.55 | 9.10 | 15.10 | 26.04 |
| 57 | FIN | 4.06 | 7.07 | 18.10 | 5.23 | 7.95 | 18.73 |
| 58 | FJI | 4.82 | 7.06 | 25.14 | 7.05 | 11.75 | 31.37 |
| 59 | FRA | 4.50 | 8.38 | 24.56 | 5.70 | 10.75 | 22.93 |
| 60 | GAB | 4.38 | 7.27 | 12.11 | 7.58 | 12.14 | 19.17 |
| 61 | GBR | 2.88 | 4.85 | 12.05 | 4.33 | 7.18 | 14.12 |
| 62 | GHA | 6.08 | 9.09 | 17.87 | 8.14 | 14.14 | 24.95 |
| 63 | GIN | 6.66 | 11.65 | 25.22 | 9.65 | 16.85 | 33.26 |
| 64 | GMB | 4.79 | 7.83 | 31.14 | 8.45 | 14.51 | 34.85 |
| 65 | GNB | 4.88 | 7.26 | 25.43 | 8.31 | 13.38 | 30.09 |
| 66 | GNQ | | | | 8.24 | 13.98 | 23.39 |
| 67 | GRC | 5.18 | 8.77 | 22.75 | 6.33 | 10.68 | 23.26 |
| 68 | GRD | 7.51 | 12.56 | 20.51 | 10.78 | 17.46 | 27.19 |
| 69 | GTM | 5.48 | 7.58 | 12.34 | 6.78 | 10.14 | 18.99 |
| 70 | HKG | 6.76 | 10.69 | 43.27 | 8.66 | 14.05 | 45.36 |
| 71 | HND | 4.23 | 6.26 | 11.06 | 7.60 | 12.48 | 20.02 |
| 72 | HRV | 5.54 | 10.02 | 29.11 | 7.68 | 11.97 | 29.30 |
| 73 | HUN | 5.34 | 7.82 | 17.96 | 7.52 | 11.19 | 22.12 |
| 74 | IDN | 7.32 | 11.55 | 18.48 | 8.75 | 14.46 | 31.00 |
| 75 | IND | 5.27 | 8.97 | 16.76 | 7.44 | 12.52 | 31.12 |
| 76 | IRL | 3.67 | 6.92 | 14.03 | 4.99 | 8.22 | 14.71 |
| 77 | IRQ | 4.89 | 8.08 | 18.95 | 7.75 | 13.12 | 25.58 |
| 78 | ISL | 5.45 | 10.46 | 18.32 | 5.63 | 9.27 | 17.38 |
| 79 | ISR | 3.79 | 6.38 | 16.09 | 5.05 | 9.15 | 20.05 |
| 80 | ITA | 5.15 | 8.62 | 17.87 | 6.16 | 10.26 | 19.38 |
| 81 | JAM | 5.82 | 10.15 | 18.34 | 8.97 | 15.41 | 24.46 |
| 82 | JOR | 4.34 | 6.46 | 12.57 | 6.75 | 11.60 | 29.42 |
| 83 | JPN | 6.82 | 11.70 | 28.37 | 9.58 | 14.82 | 26.52 |
| 84 | KAZ | 5.63 | 8.92 | 13.62 | 8.40 | 14.02 | 29.74 |
| 85 | KEN | 5.64 | 9.81 | 23.03 | 7.84 | 13.09 | 24.70 |
| 86 | KGZ | 6.28 | 10.53 | 16.96 | 9.72 | 15.76 | 26.54 |
| 87 | KHM | 8.16 | 11.35 | 23.52 | 8.29 | 13.71 | 42.71 |
| 88 | KNA | 5.27 | 8.53 | 14.49 | 8.09 | 12.89 | 21.48 |
| 89 | KOR | 7.81 | 15.87 | 33.17 | 10.89 | 18.31 | 38.72 |
| 90 | KWT | 5.28 | 8.94 | 18.61 | 7.67 | 13.61 | 26.85 |
| 91 | LAO | 5.81 | 9.17 | 14.68 | 8.61 | 14.78 | 33.08 |
| 92 | LBN | 5.52 | 8.59 | 23.21 | 8.09 | 14.04 | 29.31 |
| 93 | LBR | 6.31 | 9.22 | 27.24 | 7.95 | 13.36 | 29.83 |
| 94 | LCA | 7.12 | 10.27 | 21.36 | 10.03 | 17.76 | 34.69 |
| 95 | LKA | 7.36 | 11.27 | 24.31 | 10.61 | 17.77 | 35.72 |
| 96 | LSO | 5.36 | 8.65 | 12.85 | 8.17 | 13.29 | 19.51 |
| 97 | LTU | 4.43 | 8.32 | 24.64 | 6.82 | 11.12 | 24.81 |
| 98 | LUX | 4.22 | 7.92 | 18.09 | 5.28 | 9.31 | 18.85 |
| 99 | LVA | 5.10 | 8.62 | 22.79 | 7.06 | 11.03 | 22.81 |
| 100 | MAR | 5.81 | 8.79 | 29.11 | 6.21 | 11.34 | 29.43 |
| 101 | MDA | 6.45 | 11.54 | 18.44 | 8.24 | 13.24 | 20.97 |
| 102 | MDG | 5.73 | 11.40 | 22.58 | 7.15 | 13.30 | 25.86 |
| 103 | MDV | 4.38 | 6.66 | 10.65 | 7.41 | 11.19 | 16.59 |
| 104 | MEX | 4.72 | 7.21 | 27.32 | 6.82 | 10.94 | 31.15 |
| 105 | MKD | 5.56 | 8.63 | 29.38 | 7.53 | 12.57 | 32.17 |
| 106 | MLI | 4.93 | 8.83 | 18.61 | 7.07 | 12.80 | 29.24 |
| 107 | MLT | 5.97 | 11.06 | 28.35 | 7.55 | 12.43 | 26.91 |
| 108 | MNE | 6.51 | 10.17 | 26.63 | 8.02 | 13.27 | 30.41 |
| 109 | MNG | 7.37 | 10.85 | 16.47 | 11.78 | 16.82 | 28.36 |
| 110 | MOZ | 5.08 | 8.62 | 18.10 | 7.19 | 12.66 | 22.41 |
| 111 | MRT | 8.39 | 12.23 | 17.58 | 8.02 | 14.83 | 27.25 |
| 112 | MSR | | | | 10.18 | 17.22 | 30.03 |
| 113 | MUS | 7.24 | 10.92 | 26.87 | 8.72 | 13.72 | 26.80 |
| 114 | MWI | 5.06 | 7.97 | 13.83 | 7.15 | 13.49 | 24.71 |
| 115 | MYS | 5.95 | 10.02 | 20.91 | 9.07 | 13.57 | 30.38 |
| 116 | NAM | 5.52 | 7.77 | 12.76 | 8.06 | 12.18 | 18.12 |
| 117 | NER | 6.38 | 11.02 | 23.22 | 8.75 | 14.92 | 26.03 |
| 118 | NGA | 6.36 | 8.72 | 39.78 | 7.58 | 14.09 | 41.97 |
| 119 | NIC | 4.77 | 6.71 | 16.57 | 6.50 | 11.01 | 24.11 |
| 120 | NLD | 3.67 | 7.41 | 16.78 | 4.98 | 9.45 | 17.75 |
| 121 | NOR | 5.68 | 10.92 | 20.87 | 7.80 | 12.00 | 20.13 |
| 122 | NPL | 5.80 | 8.09 | 13.05 | 7.10 | 10.76 | 17.20 |
| 123 | NZL | 3.98 | 6.97 | 16.39 | 6.05 | 9.34 | 17.16 |
| 124 | OMN | 5.74 | 9.42 | 18.13 | 7.26 | 12.29 | 26.39 |
| 125 | PAK | 5.26 | 7.98 | 14.23 | 6.32 | 11.65 | 24.02 |
| 126 | PAN | 4.67 | 8.20 | 13.97 | 7.84 | 13.20 | 20.69 |
| 127 | PER | 4.45 | 9.15 | 16.82 | 6.12 | 10.50 | 18.90 |
| 128 | PHL | 5.79 | 9.33 | 16.20 | 8.80 | 14.15 | 34.98 |
| 129 | POL | 4.34 | 7.38 | 29.00 | 7.10 | 12.07 | 32.51 |
| 130 | PRT | 5.02 | 9.54 | 27.79 | 6.11 | 9.91 | 26.96 |
| 131 | PRY | 5.28 | 7.38 | 16.55 | 7.91 | 13.73 | 24.28 |
| 132 | PSE | | | | 6.65 | 11.42 | 24.94 |
| 133 | QAT | 3.60 | 6.04 | 12.14 | 5.27 | 8.85 | 17.19 |
| 134 | ROU | 4.72 | 7.07 | 24.13 | 6.86 | 10.86 | 27.04 |
| 135 | RUS | 5.10 | 10.09 | 18.12 | 7.48 | 11.96 | 27.06 |
| 136 | RWA | 5.04 | 8.52 | 13.80 | 6.22 | 11.37 | 20.18 |
| 137 | SAU | 5.11 | 7.97 | 17.29 | 7.84 | 13.64 | 27.81 |
| 138 | SDN | 4.42 | 6.55 | 11.73 | 6.68 | 11.79 | 23.56 |
| 139 | SEN | 5.39 | 8.57 | 22.00 | 7.37 | 12.67 | 23.68 |
| 140 | SGP | | | | 6.58 | 10.76 | 18.76 |
| 141 | SLE | 5.77 | 9.65 | 16.96 | 8.34 | 14.57 | 26.68 |
| 142 | SOM | 6.37 | 9.85 | 15.04 | 7.41 | 11.87 | 17.08 |
| 143 | SRB | 6.33 | 9.76 | 25.05 | 8.94 | 14.63 | 30.09 |
| 144 | SSD | 7.47 | 9.70 | 27.00 | 8.19 | 12.72 | 29.10 |
| 145 | STP | 6.00 | 9.41 | 14.28 | 8.97 | 14.88 | 27.79 |
| 146 | SUR | 6.64 | 10.19 | 17.10 | 11.87 | 19.10 | 30.54 |
| 147 | SVK | 5.56 | 9.12 | 20.44 | 6.86 | 10.97 | 20.88 |
| 148 | SVN | 5.16 | 7.76 | 19.83 | 6.35 | 9.94 | 21.10 |
| 149 | SWE | 4.38 | 8.59 | 18.46 | 6.37 | 10.87 | 20.37 |
| 150 | SWZ | 5.21 | 7.41 | 23.12 | 6.61 | 11.54 | 27.59 |
| 151 | SYC | 5.77 | 9.12 | 16.02 | 8.36 | 13.59 | 21.43 |
| 152 | SYR | 8.01 | 12.42 | 24.60 | 9.31 | 15.04 | 28.92 |
| 153 | TCD | 4.05 | 6.71 | 11.06 | 7.21 | 12.20 | 21.15 |
| 154 | TGO | 4.82 | 7.80 | 13.48 | 7.12 | 12.39 | 22.12 |
| 155 | THA | 7.47 | 11.97 | 33.22 | 9.94 | 16.20 | 38.42 |
| 156 | TJK | 6.79 | 11.93 | 19.87 | 9.60 | 16.02 | 30.23 |
| 157 | TTO | 5.75 | 10.53 | 17.50 | 9.65 | 15.40 | 22.24 |
| 158 | TUN | 6.46 | 10.56 | 46.02 | 7.22 | 14.52 | 46.00 |
| 159 | TUR | 5.09 | 7.54 | 29.81 | 7.08 | 12.55 | 33.43 |
| 160 | TWN | 7.87 | 11.02 | 45.02 | 9.97 | 15.78 | 48.71 |
| 161 | TZA | 3.96 | 6.89 | 12.27 | 6.10 | 11.90 | 21.16 |
| 162 | UGA | 5.53 | 9.41 | 16.85 | 6.26 | 11.24 | 19.84 |
| 163 | URY | 4.23 | 6.92 | 11.50 | 6.72 | 10.97 | 17.34 |
| 164 | USA | 3.45 | 5.59 | 13.46 | 4.77 | 7.79 | 16.07 |
| 165 | UZB | 7.19 | 11.11 | 20.36 | 11.41 | 19.23 | 37.44 |
| 166 | VCT | 6.21 | 10.92 | 18.15 | 11.40 | 18.14 | 26.53 |
| 167 | VGB | | | | 8.73 | 12.30 | 17.07 |
| 168 | VNM | 6.12 | 9.75 | 50.91 | 8.46 | 14.39 | 55.63 |
| 169 | ZAF | 3.72 | 5.53 | 9.84 | 5.85 | 9.80 | 20.85 |
| 170 | ZMB | 5.07 | 7.74 | 21.80 | 6.56 | 11.67 | 25.19 |
| 171 | ZWE | 4.33 | 6.95 | 10.72 | 7.02 | 12.01 | 24.54 |
| **Global average** | | **5.47** | **8.90** | **20.79** | **7.60** | **12.71** | **26.09** |

Suppl. Table 4. Summary statistics for GHG emissions using Monte Carlo simulation in each country

| No | Country | Diet 4: All food items, weighted by use GHG Emissions (kg CO2e) | | | Diet 5: All food items, weighted equally GHG Emissions (kg CO2e) | | |
|---|---|---|---|---|---|---|---|
| | | P2.5 | P50 | P97.5 | P2.5 | P50 | P97.5 |
| 1 | ABW | | | | 1.37 | 2.67 | 5.69 |
| 2 | AGO | 0.96 | 2.18 | 4.31 | 1.23 | 2.50 | 5.51 |
| 3 | AIA | | | | 1.40 | 2.81 | 6.02 |
| 4 | ALB | 1.13 | 2.30 | 5.37 | 1.40 | 2.76 | 5.44 |
| 5 | ARE | 1.00 | 2.14 | 4.70 | 1.34 | 2.87 | 6.37 |
| 6 | ARG | 1.23 | 3.21 | 6.54 | 1.31 | 2.68 | 5.52 |
| 7 | ARM | 1.25 | 2.59 | 6.12 | 1.39 | 2.40 | 5.55 |
| 8 | ATG | 0.86 | 1.56 | 3.62 | 1.32 | 2.59 | 6.09 |
| 9 | AUS | 1.05 | 1.87 | 4.78 | 1.35 | 2.69 | 5.35 |
| 10 | AUT | 0.96 | 1.64 | 3.86 | 1.31 | 2.56 | 5.14 |
| 11 | AZE | 1.32 | 3.14 | 6.51 | 1.44 | 2.55 | 5.80 |
| 12 | BDI | 0.89 | 2.14 | 4.83 | 1.21 | 2.53 | 5.64 |
| 13 | BEL | 0.87 | 1.61 | 3.99 | 1.27 | 2.67 | 5.45 |
| 14 | BEN | 1.21 | 2.43 | 5.43 | 1.30 | 2.62 | 7.38 |
| 15 | BFA | 1.05 | 2.16 | 4.27 | 1.26 | 2.58 | 7.68 |
| 16 | BGD | 1.28 | 2.64 | 4.79 | 1.46 | 3.27 | 14.48 |
| 17 | BGR | 1.04 | 1.57 | 3.74 | 1.32 | 2.46 | 5.05 |
| 18 | BHR | 0.97 | 2.47 | 5.02 | 1.37 | 2.84 | 6.53 |
| 19 | BIH | 1.09 | 2.41 | 5.45 | 1.40 | 2.71 | 5.51 |
| 20 | BLR | 1.13 | 1.99 | 4.72 | 1.48 | 2.42 | 5.49 |
| 21 | BLZ | 1.47 | 2.17 | 4.64 | 1.44 | 2.99 | 6.75 |
| 22 | BMU | | | | 1.36 | 2.84 | 5.95 |
| 23 | BOL | 1.10 | 2.32 | 6.14 | 1.27 | 2.26 | 5.36 |
| 24 | BON | | | | 1.38 | 2.53 | 5.57 |
| 25 | BRA | 1.16 | 1.90 | 5.05 | 1.42 | 2.76 | 5.50 |
| 26 | BRN | | | | 1.42 | 2.82 | 7.34 |
| 27 | BTN | 1.28 | 1.98 | 3.42 | 1.31 | 2.22 | 5.04 |
| 28 | BWA | 1.07 | 1.55 | 4.76 | 1.30 | 2.69 | 6.21 |
| 29 | CAF | 1.26 | 3.46 | 6.16 | 1.31 | 2.64 | 7.53 |
| 30 | CAN | 1.00 | 1.75 | 4.10 | 1.35 | 2.45 | 5.03 |
| 31 | CHE | 1.17 | 1.68 | 3.85 | 1.38 | 2.59 | 5.40 |
| 32 | CHL | 1.08 | 1.71 | 4.42 | 1.31 | 2.08 | 5.29 |
| 33 | CHN | 1.42 | 2.95 | 7.51 | 1.50 | 3.63 | 9.11 |
| 34 | CIV | 0.97 | 2.26 | 3.99 | 1.27 | 2.61 | 7.46 |
| 35 | CMR | 0.97 | 2.26 | 4.93 | 1.31 | 2.62 | 7.92 |
| 36 | COD | 0.83 | 2.12 | 4.04 | 1.33 | 2.71 | 7.47 |
| 37 | COG | 0.89 | 1.83 | 3.59 | 1.27 | 2.61 | 7.29 |
| 38 | COL | 1.21 | 2.42 | 5.32 | 1.39 | 2.71 | 5.60 |
| 39 | COM | 1.04 | 1.99 | 3.90 | 1.28 | 2.64 | 7.42 |
| 40 | CPV | 1.07 | 1.64 | 4.05 | 1.28 | 2.39 | 7.31 |
| 41 | CRI | 0.99 | 1.59 | 3.98 | 1.31 | 2.14 | 4.97 |
| 42 | CUW | | | | 1.41 | 2.75 | 6.25 |
| 43 | CYM | | | | 1.40 | 2.24 | 5.37 |
| 44 | CYP | 0.99 | 1.67 | 4.07 | 1.31 | 2.50 | 5.34 |
| 45 | CZE | 0.94 | 1.58 | 3.93 | 1.31 | 2.55 | 5.31 |
| 46 | DEU | 1.02 | 1.48 | 3.91 | 1.34 | 2.53 | 5.32 |
| 47 | DJI | 1.39 | 3.26 | 7.65 | 1.38 | 2.68 | 7.48 |
| 48 | DMA | 1.07 | 1.69 | 4.17 | 1.37 | 2.59 | 5.67 |
| 49 | DNK | 1.05 | 1.90 | 4.36 | 1.28 | 2.57 | 5.08 |
| 50 | DOM | 1.24 | 1.89 | 4.41 | 1.38 | 2.56 | 5.99 |
| 51 | DZA | 1.04 | 2.59 | 6.04 | 1.32 | 2.58 | 6.04 |
| 52 | ECU | 1.31 | 2.51 | 5.44 | 1.40 | 2.76 | 5.63 |
| 53 | EGY | 0.89 | 2.30 | 4.75 | 1.47 | 3.24 | 7.32 |
| 54 | ESP | 0.91 | 1.56 | 3.77 | 1.36 | 2.61 | 5.36 |
| 55 | EST | 1.09 | 1.87 | 4.44 | 1.37 | 2.62 | 5.46 |
| 56 | ETH | 1.20 | 3.48 | 7.81 | 1.30 | 2.66 | 7.49 |
| 57 | FIN | 1.03 | 1.74 | 4.29 | 1.26 | 2.04 | 4.51 |
| 58 | FJI | 1.03 | 2.19 | 6.16 | 1.31 | 2.59 | 6.87 |
| 59 | FRA | 0.91 | 1.67 | 4.19 | 1.26 | 2.48 | 5.29 |
| 60 | GAB | 0.82 | 1.81 | 3.48 | 1.23 | 2.45 | 7.40 |
| 61 | GBR | 0.97 | 1.48 | 3.88 | 1.33 | 2.50 | 5.19 |
| 62 | GHA | 1.02 | 2.16 | 3.95 | 1.27 | 2.59 | 7.39 |
| 63 | GIN | 1.08 | 2.41 | 5.10 | 1.34 | 2.67 | 7.55 |
| 64 | GMB | 1.07 | 2.09 | 3.79 | 1.29 | 2.55 | 8.06 |
| 65 | GNB | 1.26 | 2.49 | 6.28 | 1.31 | 2.69 | 7.35 |
| 66 | GNQ | | | | 1.38 | 2.67 | 6.60 |
| 67 | GRC | 1.10 | 1.82 | 4.29 | 1.33 | 2.56 | 5.27 |
| 68 | GRD | 1.12 | 1.77 | 3.83 | 1.38 | 2.66 | 5.96 |
| 69 | GTM | 1.22 | 1.70 | 4.09 | 1.47 | 2.53 | 5.37 |
| 70 | HKG | 1.58 | 3.05 | 7.44 | 1.60 | 3.09 | 7.49 |
| 71 | HND | 1.17 | 1.76 | 4.07 | 1.48 | 2.78 | 5.67 |
| 72 | HRV | 0.98 | 2.21 | 4.83 | 1.34 | 2.53 | 4.88 |
| 73 | HUN | 1.05 | 1.49 | 2.78 | 1.26 | 2.33 | 5.03 |
| 74 | IDN | 1.50 | 3.03 | 6.39 | 1.45 | 3.21 | 7.76 |
| 75 | IND | 1.20 | 2.21 | 5.20 | 1.52 | 3.49 | 13.30 |
| 76 | IRL | 1.02 | 2.25 | 5.33 | 1.30 | 2.49 | 5.30 |
| 77 | IRQ | 1.01 | 1.83 | 4.13 | 1.35 | 2.96 | 6.62 |
| 78 | ISL | 1.19 | 2.34 | 4.76 | 1.36 | 2.60 | 5.42 |
| 79 | ISR | 1.08 | 2.61 | 5.43 | 1.29 | 2.59 | 5.37 |
| 80 | ITA | 1.21 | 2.63 | 5.22 | 1.47 | 2.69 | 5.26 |
| 81 | JAM | 1.01 | 1.80 | 4.16 | 1.38 | 2.74 | 6.18 |
| 82 | JOR | 1.07 | 1.82 | 4.56 | 1.37 | 3.19 | 6.91 |
| 83 | JPN | 1.43 | 1.99 | 4.20 | 1.52 | 2.67 | 5.07 |
| 84 | KAZ | 1.29 | 3.03 | 6.80 | 1.39 | 2.36 | 5.41 |
| 85 | KEN | 1.14 | 2.46 | 6.38 | 1.31 | 2.64 | 7.32 |
| 86 | KGZ | 1.25 | 3.49 | 6.72 | 1.45 | 2.49 | 5.80 |
| 87 | KHM | 1.42 | 2.77 | 6.13 | 1.55 | 3.19 | 13.89 |
| 88 | KNA | 1.09 | 1.86 | 3.87 | 1.41 | 2.68 | 5.50 |
| 89 | KOR | 1.27 | 2.54 | 4.94 | 1.42 | 2.68 | 5.36 |
| 90 | KWT | 1.10 | 2.13 | 5.18 | 1.35 | 3.20 | 6.99 |
| 91 | LAO | 1.29 | 2.48 | 5.53 | 1.44 | 3.14 | 8.99 |
| 92 | LBN | 1.07 | 2.22 | 5.07 | 1.32 | 2.83 | 6.65 |
| 93 | LBR | 1.32 | 2.03 | 4.05 | 1.32 | 2.53 | 5.66 |
| 94 | LCA | 0.86 | 1.47 | 3.74 | 1.43 | 2.58 | 5.41 |
| 95 | LKA | 1.30 | 2.51 | 5.72 | 1.47 | 3.03 | 7.87 |
| 96 | LSO | 0.98 | 1.86 | 6.97 | 1.37 | 2.80 | 7.75 |
| 97 | LTU | 0.92 | 1.67 | 3.70 | 1.34 | 2.51 | 5.15 |
| 98 | LUX | 1.12 | 2.55 | 5.43 | 1.36 | 2.62 | 5.32 |
| 99 | LVA | 0.90 | 1.76 | 4.10 | 1.34 | 2.56 | 5.43 |
| 100 | MAR | 0.86 | 2.17 | 6.82 | 1.31 | 2.64 | 7.32 |
| 101 | MDA | 1.11 | 1.73 | 3.09 | 1.39 | 2.45 | 5.47 |
| 102 | MDG | 1.30 | 2.49 | 4.83 | 1.32 | 2.68 | 7.55 |
| 103 | MDV | 0.91 | 1.69 | 3.36 | 1.40 | 2.11 | 5.79 |
| 104 | MEX | 1.17 | 1.75 | 3.96 | 1.39 | 2.72 | 5.27 |
| 105 | MKD | 1.11 | 2.02 | 5.13 | 1.32 | 2.54 | 5.34 |
| 106 | MLI | 1.00 | 2.17 | 4.80 | 1.31 | 2.61 | 7.05 |
| 107 | MLT | 0.98 | 1.66 | 3.96 | 1.32 | 2.54 | 5.31 |
| 108 | MNE | 1.30 | 2.03 | 4.97 | 1.27 | 2.63 | 5.29 |
| 109 | MNG | 1.63 | 4.64 | 7.65 | 1.60 | 3.04 | 6.55 |
| 110 | MOZ | 1.03 | 1.67 | 3.55 | 1.29 | 2.51 | 7.49 |
| 111 | MRT | 1.30 | 2.77 | 8.72 | 1.35 | 2.91 | 8.64 |
| 112 | MSR | | | | 1.46 | 2.75 | 5.80 |
| 113 | MUS | 0.96 | 1.73 | 4.23 | 1.29 | 2.56 | 6.42 |
| 114 | MWI | 1.01 | 1.71 | 4.13 | 1.27 | 2.46 | 7.41 |
| 115 | MYS | 1.36 | 2.60 | 6.11 | 1.47 | 3.21 | 13.94 |
| 116 | NAM | 0.80 | 1.58 | 4.65 | 1.29 | 2.56 | 6.10 |
| 117 | NER | 1.39 | 2.83 | 6.88 | 1.29 | 2.62 | 7.47 |
| 118 | NGA | 0.99 | 2.24 | 4.80 | 1.29 | 2.66 | 7.81 |
| 119 | NIC | 1.22 | 1.72 | 4.12 | 1.44 | 2.71 | 5.36 |
| 120 | NLD | 1.13 | 2.13 | 5.65 | 1.32 | 2.50 | 5.39 |
| 121 | NOR | 1.09 | 1.78 | 4.28 | 1.38 | 2.63 | 5.69 |
| 122 | NPL | 1.10 | 1.77 | 13.72 | 1.18 | 2.02 | 13.54 |
| 123 | NZL | 0.96 | 1.71 | 4.62 | 1.41 | 2.60 | 5.50 |
| 124 | OMN | 1.17 | 2.44 | 4.85 | 1.34 | 2.92 | 6.87 |
| 125 | PAK | 1.23 | 3.05 | 7.27 | 1.54 | 3.43 | 13.29 |
| 126 | PAN | 1.21 | 1.86 | 4.14 | 1.41 | 2.45 | 5.23 |
| 127 | PER | 1.26 | 2.09 | 4.40 | 1.41 | 2.66 | 5.72 |
| 128 | PHL | 1.36 | 2.35 | 5.23 | 1.52 | 3.28 | 13.69 |
| 129 | POL | 0.93 | 1.49 | 2.98 | 1.33 | 2.53 | 5.34 |
| 130 | PRT | 0.91 | 2.12 | 4.96 | 1.34 | 2.49 | 5.34 |
| 131 | PRY | 1.05 | 1.62 | 4.21 | 1.38 | 2.67 | 5.65 |
| 132 | PSE | | | | 1.41 | 3.19 | 6.79 |
| 133 | QAT | 1.05 | 2.33 | 5.25 | 1.46 | 3.33 | 7.05 |
| 134 | ROU | 1.02 | 1.49 | 3.89 | 1.35 | 2.54 | 5.20 |
| 135 | RUS | 1.03 | 1.90 | 4.19 | 1.42 | 2.43 | 5.15 |
| 136 | RWA | 0.89 | 2.26 | 5.42 | 1.24 | 2.53 | 7.39 |
| 137 | SAU | 1.04 | 1.90 | 4.32 | 1.34 | 2.87 | 6.57 |
| 138 | SDN | 1.15 | 3.61 | 8.72 | 1.21 | 2.91 | 7.98 |
| 139 | SEN | 1.19 | 2.54 | 6.29 | 1.29 | 2.62 | 7.46 |
| 140 | SGP | | | | 1.45 | 2.85 | 6.35 |
| 141 | SLE | 1.15 | 2.36 | 3.94 | 1.29 | 2.67 | 7.65 |
| 142 | SOM | 1.52 | 3.96 | 9.50 | 1.46 | 3.63 | 8.50 |
| 143 | SRB | 1.07 | 2.17 | 4.83 | 1.32 | 2.59 | 5.23 |
| 144 | SSD | 1.10 | 2.38 | 6.36 | 1.29 | 2.61 | 7.61 |
| 145 | STP | 0.90 | 1.59 | 3.74 | 1.27 | 2.21 | 6.00 |
| 146 | SUR | 1.23 | 1.89 | 4.21 | 1.38 | 2.76 | 6.13 |
| 147 | SVK | 1.03 | 1.68 | 3.72 | 1.36 | 2.54 | 5.16 |
| 148 | SVN | 1.03 | 1.63 | 3.73 | 1.34 | 2.45 | 5.08 |
| 149 | SWE | 0.94 | 1.95 | 4.29 | 1.33 | 2.71 | 5.31 |
| 150 | SWZ | 1.15 | 2.87 | 6.06 | 1.33 | 2.41 | 5.50 |
| 151 | SYC | 0.92 | 1.67 | 4.04 | 1.31 | 2.57 | 6.27 |
| 152 | SYR | 1.20 | 3.63 | 6.48 | 1.39 | 3.24 | 6.45 |
| 153 | TCD | 1.67 | 4.19 | 9.20 | 1.30 | 2.57 | 7.43 |
| 154 | TGO | 0.85 | 1.99 | 4.03 | 1.24 | 2.63 | 7.26 |
| 155 | THA | 1.51 | 3.03 | 7.16 | 1.54 | 3.09 | 7.51 |
| 156 | TJK | 1.45 | 3.80 | 6.40 | 1.40 | 2.47 | 5.40 |
| 157 | TTO | 1.02 | 1.76 | 3.91 | 1.42 | 2.84 | 5.99 |
| 158 | TUN | 1.00 | 2.14 | 7.14 | 1.36 | 2.85 | 8.26 |
| 159 | TUR | 1.08 | 1.57 | 3.84 | 1.25 | 2.17 | 4.35 |
| 160 | TWN | 1.60 | 2.74 | 7.32 | 1.43 | 2.71 | 7.40 |
| 161 | TZA | 0.98 | 2.42 | 5.15 | 1.27 | 2.63 | 7.34 |
| 162 | UGA | 0.81 | 1.72 | 3.97 | 1.29 | 2.32 | 7.34 |
| 163 | URY | 1.04 | 1.88 | 4.64 | 1.23 | 2.41 | 5.29 |
| 164 | USA | 1.07 | 1.65 | 4.16 | 1.33 | 2.53 | 5.30 |
| 165 | UZB | 1.19 | 3.50 | 6.52 | 1.43 | 2.53 | 5.69 |
| 166 | VCT | 1.02 | 1.65 | 3.25 | 1.41 | 2.71 | 6.01 |
| 167 | VGB | | | | 1.37 | 2.84 | 6.86 |
| 168 | VNM | 1.35 | 2.54 | 6.70 | 1.40 | 2.75 | 13.34 |
| 169 | ZAF | 1.10 | 2.30 | 6.31 | 1.31 | 2.46 | 7.80 |
| 170 | ZMB | 0.82 | 2.22 | 4.83 | 1.29 | 2.57 | 7.28 |
| 171 | ZWE | 1.47 | 3.70 | 6.27 | 1.32 | 2.69 | 7.43 |
| **Global average** | | **1.12** | **2.21** | **5.06** | **1.36** | **2.67** | **6.60** |

Suppl. Table 5. Summary statistics for monetary cost by food group (2021 PPP dollars per person per day)

| Diet | Obs | Mean | SD | P25 | P50 | P75 | t-test, compared to case 1, p-value (diff ≠ 0) |
|---|---|---|---|---|---|---|---|
| **Animal-source Food** | | | | | | | |
| 1 | 171 | 1.00 | 0.28 | 0.80 | 0.99 | 1.19 | na |
| 2 | 171 | 2.52 | 1.04 | 1.84 | 2.29 | 2.91 | 0.00 |
| 3 | 159 | 1.54 | 0.59 | 1.13 | 1.38 | 1.86 | 0.00 |
| 4 | 159,000 | 4.23 | 4.41 | 1.96 | 2.94 | 4.71 | 0.00 |
| 5 | 171,000 | 4.48 | 4.15 | 2.19 | 3.27 | 5.14 | 0.00 |
| **Vegetables** | | | | | | | |
| 1 | 171 | 0.77 | 0.30 | 0.54 | 0.76 | 0.95 | na |
| 2 | 171 | 0.97 | 0.35 | 0.72 | 0.89 | 1.16 | 0.00 |
| 3 | 159 | 1.47 | 0.65 | 1.01 | 1.28 | 1.73 | 0.00 |
| 4 | 159,000 | 1.65 | 0.95 | 0.98 | 1.44 | 2.08 | 0.00 |
| 5 | 171,000 | 2.23 | 1.17 | 1.42 | 1.97 | 2.74 | 0.00 |
| **Starchy Staples** | | | | | | | |
| 1 | 171 | 0.58 | 0.20 | 0.46 | 0.58 | 0.68 | na |
| 2 | 171 | 1.15 | 0.59 | 0.71 | 0.99 | 1.44 | 0.00 |
| 3 | 159 | 1.06 | 0.76 | 0.60 | 0.88 | 1.27 | 0.00 |
| 4 | 159,000 | 1.33 | 1.11 | 0.69 | 1.00 | 1.55 | 0.00 |
| 5 | 171,000 | 3.05 | 2.30 | 1.51 | 2.37 | 3.87 | 0.00 |
| **Fruits** | | | | | | | |
| 1 | 171 | 0.77 | 0.29 | 0.60 | 0.74 | 0.88 | na |
| 2 | 171 | 1.57 | 0.79 | 1.03 | 1.39 | 1.87 | 0.00 |
| 3 | 159 | 1.09 | 0.53 | 0.70 | 0.91 | 1.32 | 0.00 |
| 4 | 159,000 | 1.16 | 0.67 | 0.72 | 0.96 | 1.42 | 0.00 |
| 5 | 171,000 | 1.83 | 1.00 | 1.14 | 1.59 | 2.27 | 0.00 |
| **Legumes, nuts and seeds** | | | | | | | |
| 1 | 171 | 0.39 | 0.15 | 0.31 | 0.36 | 0.44 | na |
| 2 | 171 | 0.51 | 0.27 | 0.33 | 0.42 | 0.58 | 0.00 |
| 3 | 159 | 0.88 | 1.21 | 0.36 | 0.51 | 0.83 | 0.00 |
| 4 | 159,000 | 1.15 | 1.30 | 0.42 | 0.64 | 1.16 | 0.00 |
| 5 | 171,000 | 1.44 | 1.51 | 0.48 | 0.81 | 1.80 | 0.00 |
| **Oils and fats** | | | | | | | |
| 1 | 171 | 0.17 | 0.08 | 0.12 | 0.16 | 0.21 | na |
| 2 | 171 | 0.23 | 0.12 | 0.14 | 0.22 | 0.30 | 0.00 |
| 3 | 159 | 0.30 | 0.27 | 0.14 | 0.22 | 0.33 | 0.00 |
| 4 | 159,000 | 0.44 | 0.49 | 0.18 | 0.26 | 0.46 | 0.00 |
| 5 | 171,000 | 0.70 | 0.81 | 0.22 | 0.39 | 0.90 | 0.00 |

Suppl. Table 6. Summary statistics for GHG emissions by food group (kg CO2e per person per day)

| Diet | Obs | Mean | SD | P25 | P50 | P75 | t-test, compared to case 1, p-value (diff ≠ 0) |
|---|---|---|---|---|---|---|---|
| **Animal-source Food** | | | | | | | |
| 1 | 171 | 0.97 | 0.58 | 0.59 | 0.72 | 1.43 | na |
| 2 | 171 | 0.26 | 0.09 | 0.23 | 0.23 | 0.31 | 0.00 |
| 3 | 159 | 1.04 | 0.59 | 0.58 | 0.82 | 1.47 | 0.16 |
| 4 | 159,000 | 1.54 | 1.24 | 0.66 | 1.20 | 1.96 | 0.00 |
| 5 | 171,000 | 1.87 | 1.45 | 0.76 | 1.54 | 2.85 | 0.00 |
| **Vegetables** | | | | | | | |
| 1 | 171 | 0.10 | 0.03 | 0.08 | 0.11 | 0.11 | na |
| 2 | 171 | 0.08 | 0.01 | 0.07 | 0.08 | 0.08 | 0.00 |
| 3 | 159 | 0.18 | 0.07 | 0.14 | 0.17 | 0.22 | 0.00 |
| 4 | 159,000 | 0.19 | 0.09 | 0.13 | 0.19 | 0.25 | 0.00 |
| 5 | 171,000 | 0.23 | 0.10 | 0.17 | 0.22 | 0.28 | 0.00 |
| **Starchy Staples** | | | | | | | |
| 1 | 171 | 0.36 | 0.12 | 0.27 | 0.43 | 0.43 | na |
| 2 | 171 | 0.18 | 0.02 | 0.16 | 0.17 | 0.18 | 0.00 |
| 3 | 159 | 0.35 | 0.11 | 0.25 | 0.32 | 0.44 | 0.28 |
| 4 | 159,000 | 0.38 | 0.20 | 0.17 | 0.33 | 0.49 | 0.13 |
| 5 | 171,000 | 0.52 | 0.26 | 0.37 | 0.45 | 0.56 | 0.00 |
| **Fruits** | | | | | | | |
| 1 | 171 | 0.12 | 0.05 | 0.09 | 0.11 | 0.12 | na |
| 2 | 171 | 0.07 | 0.02 | 0.07 | 0.08 | 0.08 | 0.00 |
| 3 | 159 | 0.12 | 0.05 | 0.09 | 0.11 | 0.12 | 0.67 |
| 4 | 159,000 | 0.12 | 0.06 | 0.09 | 0.10 | 0.13 | 0.72 |
| 5 | 171,000 | 0.16 | 0.07 | 0.11 | 0.14 | 0.21 | 0.00 |
| **Legumes, nuts and seeds** | | | | | | | |
| 1 | 171 | 0.04 | 0.01 | 0.04 | 0.04 | 0.04 | na |
| 2 | 171 | 0.04 | 0.00 | 0.04 | 0.04 | 0.04 | 0.00 |
| 3 | 159 | 0.10 | 0.18 | 0.04 | 0.04 | 0.06 | 0.00 |
| 4 | 159,000 | 0.10 | 0.15 | 0.04 | 0.04 | 0.06 | 0.00 |
| 5 | 171,000 | 0.11 | 0.14 | 0.04 | 0.05 | 0.10 | 0.00 |
| **Oils and fats** | | | | | | | |
| 1 | 171 | 0.06 | 0.02 | 0.03 | 0.06 | 0.08 | na |
| 2 | 171 | 0.04 | 0.01 | 0.03 | 0.03 | 0.03 | 0.00 |
| 3 | 159 | 0.08 | 0.07 | 0.03 | 0.06 | 0.08 | 0.00 |
| 4 | 159,000 | 0.11 | 0.12 | 0.06 | 0.07 | 0.10 | 0.00 |
| 5 | 171,000 | 0.13 | 0.14 | 0.07 | 0.08 | 0.11 | 0.00 |

Suppl. Table 7. Summary statistics for nutrient adequacy of each diet (proportion of lower bounds)

| Diet | Nutrient | Obs | Mean | SD | p25 | p50 | p75 |
|---|---|---|---|---|---|---|---|
| 1 | Protein | 171 | 1.00 | 0.00 | 1.00 | 1.00 | 1.00 |
| | Calcium | 171 | 0.77 | 0.20 | 0.63 | 0.74 | 0.99 |
| | Iron | 171 | 0.97 | 0.09 | 1.00 | 1.00 | 1.00 |
| | Magnesium | 171 | 0.98 | 0.05 | 1.00 | 1.00 | 1.00 |
| | Phosphorous | 171 | 1.00 | 0.00 | 1.00 | 1.00 | 1.00 |
| | Zinc | 171 | 0.98 | 0.05 | 0.97 | 1.00 | 1.00 |
| | Copper | 171 | 1.00 | 0.00 | 1.00 | 1.00 | 1.00 |
| | Selenium | 171 | 0.98 | 0.08 | 1.00 | 1.00 | 1.00 |
| | Vitamin C | 171 | 0.86 | 0.22 | 0.66 | 1.00 | 1.00 |
| | Thiamin | 171 | 0.99 | 0.03 | 1.00 | 1.00 | 1.00 |
| | Riboflavin | 171 | 0.85 | 0.13 | 0.78 | 0.84 | 1.00 |
| | Niacin | 171 | 0.98 | 0.06 | 1.00 | 1.00 | 1.00 |
| | Vitamin B6 | 171 | 1.00 | 0.02 | 1.00 | 1.00 | 1.00 |
| | Folate | 171 | 0.99 | 0.03 | 1.00 | 1.00 | 1.00 |
| | Vitamin B12 | 171 | 0.82 | 0.19 | 0.69 | 0.87 | 1.00 |
| | Vitamin A | 171 | 0.99 | 0.08 | 1.00 | 1.00 | 1.00 |
| | **MAR** | **171** | **0.95** | **0.04** | **0.92** | **0.95** | **0.97** |
| 2 | Protein | 171 | 1.00 | 0.00 | 1.00 | 1.00 | 1.00 |
| | Calcium | 171 | 0.78 | 0.16 | 0.66 | 0.78 | 0.92 |
| | Iron | 171 | 1.00 | 0.00 | 1.00 | 1.00 | 1.00 |
| | Magnesium | 171 | 1.00 | 0.01 | 1.00 | 1.00 | 1.00 |
| | Phosphorous | 171 | 1.00 | 0.00 | 1.00 | 1.00 | 1.00 |
| | Zinc | 171 | 0.96 | 0.08 | 0.99 | 1.00 | 1.00 |
| | Copper | 171 | 1.00 | 0.00 | 1.00 | 1.00 | 1.00 |
| | Selenium | 171 | 1.00 | 0.00 | 1.00 | 1.00 | 1.00 |
| | Vitamin C | 171 | 0.51 | 0.23 | 0.33 | 0.45 | 0.72 |
| | Thiamin | 171 | 1.00 | 0.00 | 1.00 | 1.00 | 1.00 |
| | Riboflavin | 171 | 0.80 | 0.13 | 0.69 | 0.77 | 0.91 |
| | Niacin | 171 | 1.00 | 0.00 | 1.00 | 1.00 | 1.00 |
| | Vitamin B6 | 171 | 0.99 | 0.02 | 1.00 | 1.00 | 1.00 |
| | Folate | 171 | 1.00 | 0.02 | 1.00 | 1.00 | 1.00 |
| | Vitamin B12 | 171 | 1.00 | 0.05 | 1.00 | 1.00 | 1.00 |
| | Vitamin A | 171 | 1.00 | 0.06 | 1.00 | 1.00 | 1.00 |
| | **MAR** | **171** | **0.94** | **0.02** | **0.93** | **0.94** | **0.95** |
| 3 | Protein | 159 | 1.00 | 0.00 | 1.00 | 1.00 | 1.00 |
| | Calcium | 159 | 0.64 | 0.26 | 0.37 | 0.62 | 0.94 |
| | Iron | 159 | 0.96 | 0.09 | 0.97 | 1.00 | 1.00 |
| | Magnesium | 159 | 0.99 | 0.03 | 1.00 | 1.00 | 1.00 |
| | Phosphorous | 159 | 1.00 | 0.00 | 1.00 | 1.00 | 1.00 |
| | Zinc | 159 | 0.95 | 0.08 | 0.93 | 1.00 | 1.00 |
| | Copper | 159 | 1.00 | 0.00 | 1.00 | 1.00 | 1.00 |
| | Selenium | 159 | 0.97 | 0.12 | 1.00 | 1.00 | 1.00 |

| | | | | | | |
|---|---|---|---|---|---|---|
| Vitamin C | 159 | 0.96 | 0.11 | 1.00 | 1.00 | 1.00 |
| Thiamin | 159 | 0.99 | 0.03 | 1.00 | 1.00 | 1.00 |
| Riboflavin | 159 | 0.76 | 0.14 | 0.64 | 0.74 | 0.85 |
| Niacin | 159 | 1.00 | 0.02 | 1.00 | 1.00 | 1.00 |
| Vitamin B6 | 159 | 1.00 | 0.00 | 1.00 | 1.00 | 1.00 |
| Folate | 159 | 0.98 | 0.05 | 1.00 | 1.00 | 1.00 |
| Vitamin B12 | 159 | 0.88 | 0.18 | 0.82 | 1.00 | 1.00 |
| Vitamin A | 159 | 0.79 | 0.28 | 0.55 | 1.00 | 1.00 |
| **MAR** | **159** | **0.93** | **0.04** | **0.91** | **0.93** | **0.96** |

Note:

Table S7 presents the nutrient adequacy ratios (NARs) for 15 essential micronutrients and protein, along with the Mean Adequacy Ratio (MAR) over all nutrients for each diet. The table summarizes the distribution of nutrient adequacy across three diet scenarios, showing the mean, standard deviation (SD), and percentiles (p25, p50, p75).

Suppl. Table 8. Number of items in each HDB food group by region and income classification

| Name | Animal source foods | Fruits | Legumes, nuts and seeds | Fats and oils | Starchy Staples | Vegetables | Total count |
|---|---|---|---|---|---|---|---|
| **Income classification, World Bank, 2021** | | | | | | | |
| High income countries | 34 | 8 | 6 | 6 | 24 | 11 | 89 |
| Upper-middle countries | 42 | 10 | 7 | 6 | 27 | 12 | 105 |
| Lower-middle countries | 53 | 11 | 8 | 8 | 31 | 13 | 125 |
| Low income countries | 52 | 10 | 7 | 8 | 29 | 13 | 119 |
| **Region, World Bank** | | | | | | | |
| East Asia & Pacific | 51 | 10 | 7 | 7 | 31 | 15 | 122 |
| Europe & Central Asia | 41 | 9 | 6 | 6 | 26 | 11 | 100 |
| Latin America & Caribbean | 28 | 8 | 5 | 6 | 21 | 10 | 77 |
| Middle East & North Africa | 49 | 16 | 11 | 7 | 35 | 15 | 133 |
| North America | 26 | 6 | 4 | 6 | 20 | 8 | 69 |
| South Asia | 42 | 11 | 9 | 7 | 36 | 14 | 118 |
| Sub-Saharan Africa | 51 | 10 | 7 | 8 | 28 | 12 | 115 |
| **World** | **43** | **10** | **7** | **7** | **27** | **12** | **106** |

Note:

This table shows the number of food items included in the Cost of a Healthy Diet (CoHD) calculation for 2021, categorized by six Healthy Diet Basket (HDB) food groups across World Bank income classifications and regional groupings. The global average ("World") reflects all 171 countries with available data. Anguilla (AIA), Bonaire (BON), and Montserrat (MST) are excluded from income classification and regional results due to the absence of corresponding regional and income classification information in the World Bank datasets.

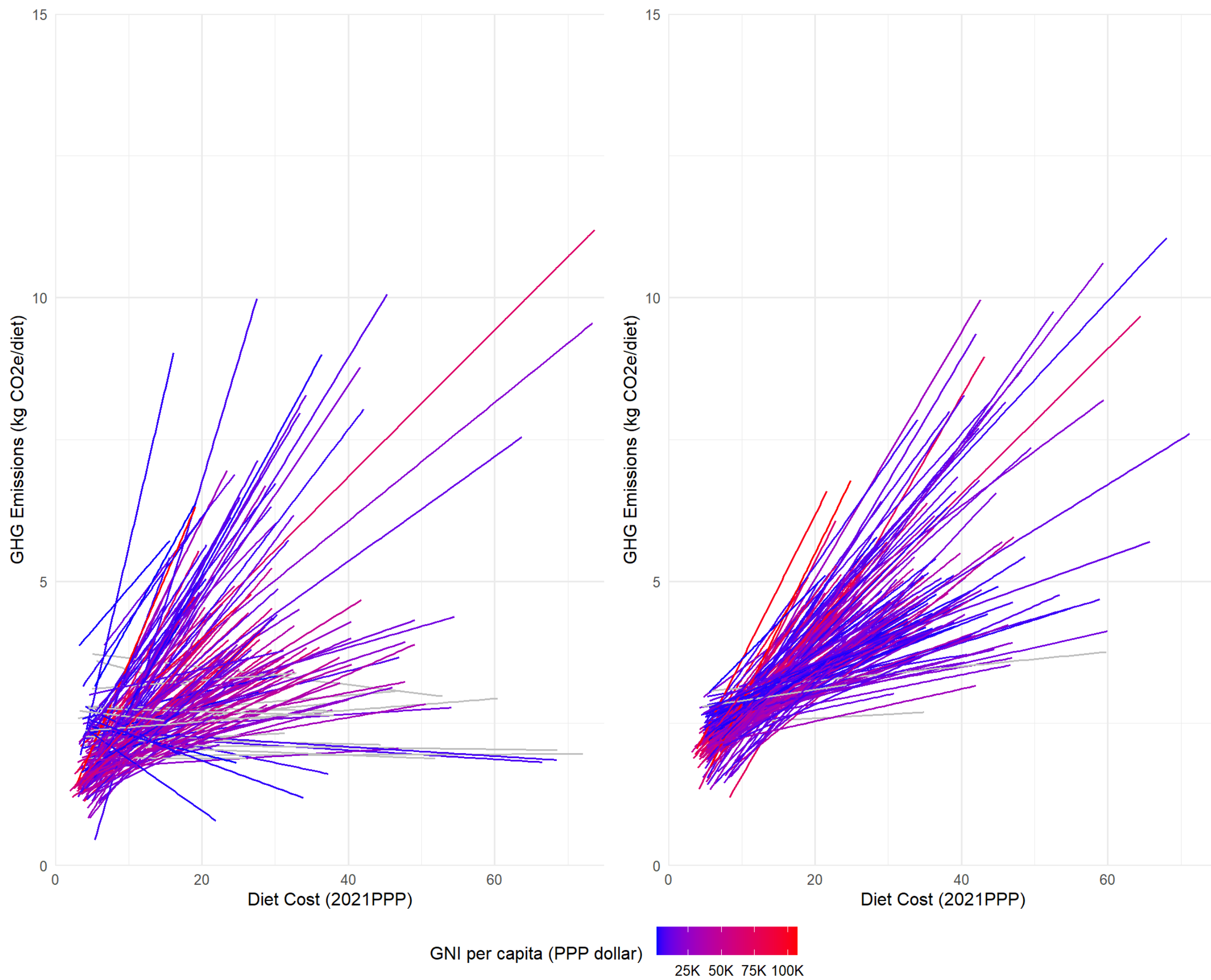

**Suppl. Fig. 1 | Relationship between diet costs and GHG emissions under Monte Carlo simulations, 2021.** The panels show the relationship between the cost of healthy diets (in 2021 PPP dollars) and associated GHG emissions (in kgs of $CO_2e$) per person per day based on estimates for 159 countries under two Monte Carlo simulation diets: (Left) weighted random selection of food items (diet 4) and (Right) random selection of food items (diet 5). Each line represents linear relationship between the two outcomes based on 1,000 Monte Carlo iterations in a country. The color gradient reflects national GNI per capita (PPP dollar), with higher incomes indicated in red and lower incomes in blue. Grey lines represent countries where the relationship between diet cost and GHG emissions is statistically insignificant ($p > 0.05$).

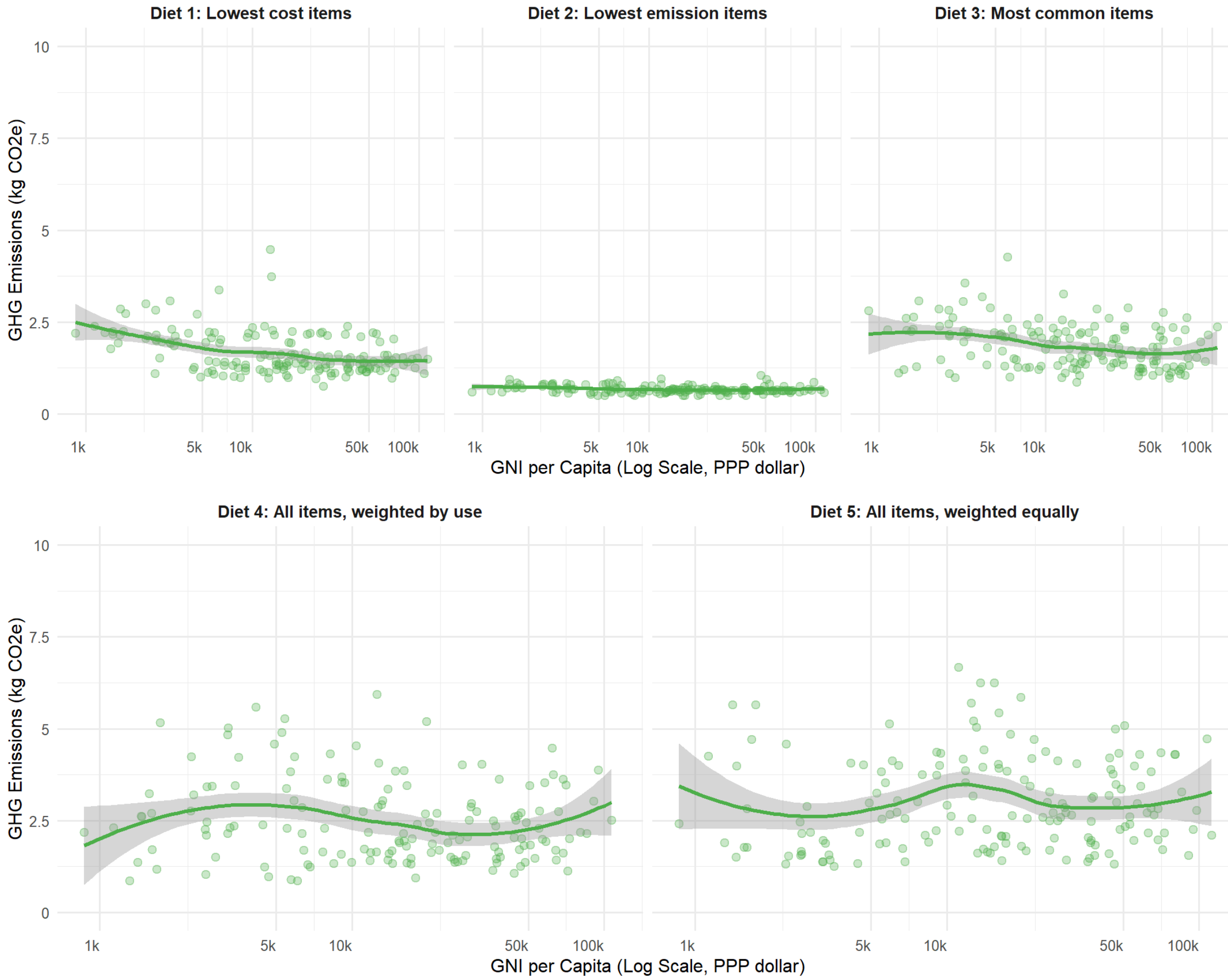

**Suppl. Fig. 2 | GHG emissions at each level of national income, by diet, 2021.** The panels show GHG emissions (in kgs of $CO_2$e per person per day at each level of Gross National Income (GNI) per capita per year (log scale, in 2021 PPP dollars). Estimates are based on data from 165 countries for diets 1, 2, and 5, and 157 countries for diets 3 and 4. Dots represent individual countries, as their national average for diets 1-3, and the median of a thousand observations for the Monte Carlo diets 4 and 5. Lines are LOESS estimates of the mean and its 95% confidence interval at each income level.

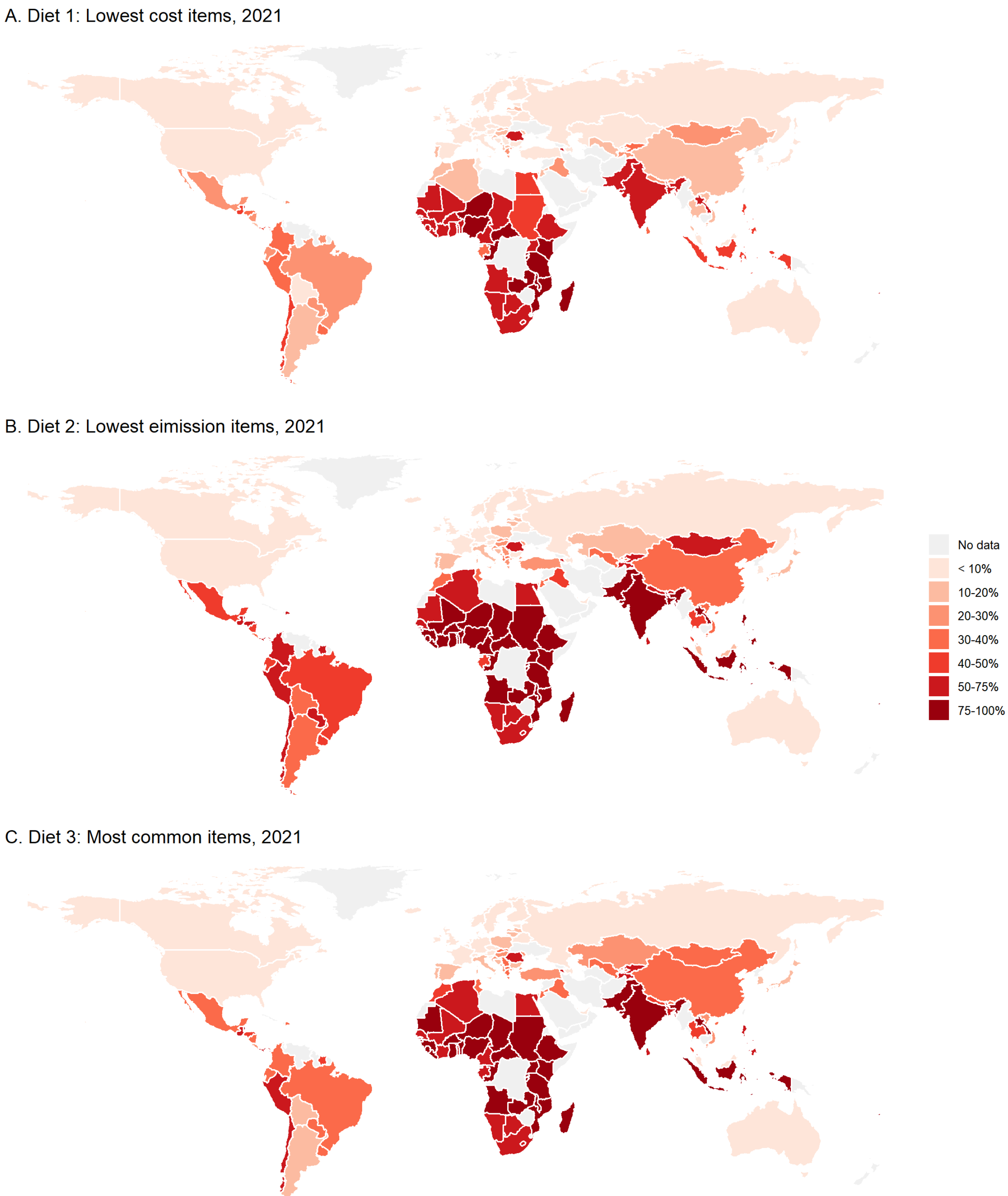

**Suppl. Fig. 3 | Prevalence of unaffordability (PUA) for a modelled diet by country, and diet, 2021.** The maps illustrate the percentage of the population who cannot afford a healthy diet under three dietary scenarios: (A) lowest cost items (diet 1), (B) lowest emission items (diet 2), and (C) most common items (diet 3). Color intensity reflects the percentage of the population unable to afford a healthy diet in each country, categorized into seven ranges: <10%, 10–20%, 20–30%, 30–40%, 40–50%, 50–75%, and 75-100%. There are 143 countries in this analysis, and countries with missing data are shown in gray. Estimates are based on affordability thresholds derived from country income classifications provided by the World Bank for the corresponding year. Further methodological details are provided in the Methods section.